\documentclass[letterpaper,11pt]{article}
\pdfoutput=1
\pdfoutput=1

\usepackage{jheppub}
\usepackage{multirow}

\usepackage{subfig}
\usepackage{xspace}
\usepackage[countmax]{subfloat}

\usepackage[normalem]{ulem}
\usepackage{amssymb}
\usepackage{amsmath}
\usepackage{cancel}
\usepackage{color}
\usepackage{braket}
\usepackage{graphicx}
\usepackage{multirow}
\usepackage{verbatim}
\usepackage{amsthm}
\usepackage{slashed}
\usepackage{wasysym}
\usepackage{simplewick}
\usepackage{mathtools}
\usepackage{soul}
\usepackage{xspace}

\newcommand{\Tau}{\mathcal{T}}

\newcount\colveccount
\newcommand*\colvec[1]{
        \global\colveccount#1
        \begin{pmatrix}
        \colvecnext
}
\def\colvecnext#1{
        #1
        \global\advance\colveccount-1
        \ifnum\colveccount>0
                \\
                \expandafter\colvecnext
        \else
                \end{pmatrix}
        \fi
}

\newcommand{\fd}[2]{\parbox{#1}{\includegraphics[width=#1]{#2}}}

\def\cB{\mathcal{B}}
\def\cC{\mathcal{C}}

\def\cL{\mathcal{L}}

\def\cN{\mathcal{N}}
\def\cO{\mathcal{O}}

\def\cP{\mathcal{P}}

\def\cY{\mathcal{Y}}

\def\tr{{\rm tr}}

\def\nn{{\nonumber}}

\newcommand{\Eq}[1]{Equation~\eqref{#1}}

\DeclareRobustCommand{\Sec}[1]{Sec.~\ref{#1}}
\DeclareRobustCommand{\Secs}[2]{Secs.~\ref{#1} and \ref{#2}}

\DeclareRobustCommand{\Tab}[1]{Table~\ref{#1}}

\DeclareRobustCommand{\Eq}[1]{Eq.~(\ref{#1})}
\DeclareRobustCommand{\Eqs}[2]{Eqs.~(\ref{#1}) and (\ref{#2})}

\def\be{\begin{equation}}
\def\ee{\end{equation}}

\newcommand{\SCETi}{\mbox{${\rm SCET}_{\rm I}$}\xspace}

\newcommand{\Sl}[1]{\slashed{#1}}

\renewcommand{\arraystretch}{1.05}
\allowdisplaybreaks[3]

\newcommand{\eq}[1]{Eq.~\eqref{eq:#1}}
\newcommand{\eqs}[2]{Eqs.~\eqref{eq:#1} and \eqref{eq:#2}}
\renewcommand{\sec}[1]{Sec.~\ref{sec:#1}}

\newcommand{\app}[1]{App.~\ref{app:#1}}

\newcommand{\df}{\mathrm{d}}

\newcommand{\Li}{\mathrm{Li}}

\newcommand\bn{{\bar n}}

\newcommand{\eps}{\epsilon}

\newcommand{\cusp}{\mathrm{cusp}}

\usepackage{marginnote}

  \newcount\hour \newcount\minute
  \hour=\time \divide \hour by 60 \minute=\time
  \newcommand{\todaytime}{\today \ -- \number\hour :\ifnum \minute<10 0\fi\number\minute}

\preprint{MIT-CTP 5156}

\title{The Soft Quark Sudakov}

\abstract{There has been recent interest in understanding the all loop structure of the subleading power soft and collinear limits, with the goal of achieving a systematic resummation of subleading power infrared logarithms. 
Most of this work has focused on subleading power corrections to soft gluon emission, whose form is strongly constrained by symmetries. In this paper we initiate a study of the all loop structure of soft fermion emission.
In $\mathcal{N}=1$ QCD we perform an operator based factorization and resummation of the associated infrared logarithms using the formalism introduced in \cite{Moult:2018jjd}, and prove that they exponentiate into a Sudakov due to their relation to soft gluon emission. We verify this result through explicit calculation to $\mathcal{O}(\alpha_s^3)$. 
We show that in QCD, this simple Sudakov exponentiation is violated by endpoint contributions proportional to $(C_A-C_F)^n$ which contribute at leading logarithmic order. Combining our $\mathcal{N}=1$ result and our calculation of the endpoint contributions to $\mathcal{O}(\alpha_s^3)$, we conjecture a result for the soft quark Sudakov in QCD, a new all orders function first appearing at subleading power, and give evidence for its universality. Our result, which is expressed in terms of combinations of cusp anomalous dimensions in different color representations, takes an intriguingly simple form and also exhibits interesting similarities to results for large-x logarithms in the off diagonal splitting functions. 
 }

\begin{document} 

\author[\psi]{Ian Moult,}
\emailAdd{imoult@slac.stanford.edu}
\author[\cB]{Iain W. Stewart,}
\emailAdd{iains@mit.edu}
\author[\cB]{Gherardo Vita,}
\emailAdd{vita@mit.edu}
\author[\chi]{and Hua Xing Zhu}
\emailAdd{zhuhx@zju.edu.cn}

\affiliation[\psi]{SLAC National Accelerator Laboratory, Stanford University, CA, 94309, USA\vspace{0.5ex}}
\affiliation[\cB]{Center for Theoretical Physics, Massachusetts Institute of Technology, Cambridge, MA 02139, USA}
\affiliation[\chi]{Department of Physics, Zhejiang University, Hangzhou, Zhejiang 310027, China\vspace{0.5ex}}

\maketitle
\section{Introduction}\label{sec:intro}

One of the most powerful approaches to understanding the all orders perturbative structure of scattering amplitudes and cross sections is to consider simplifying kinematic limits, such as the soft, collinear, or Regge limits. While much is known about the leading power behavior of these particular limits, there has recently been considerable effort to understand the all orders structure of subleading power corrections \cite{Manohar:2002fd,Beneke:2002ph,Pirjol:2002km,Beneke:2002ni,Bauer:2003mga,Hill:2004if,Lee:2004ja,Dokshitzer:2005bf,Trott:2005vw,Laenen:2008ux,Laenen:2008gt,Paz:2009ut,Soar:2009yh,
Benzke:2010js,Laenen:2010uz,Freedman:2013vya,Freedman:2014uta,Bonocore:2014wua,Larkoski:2014bxa,Bonocore:2015esa,Bonocore:2016awd,Moult:2016fqy,Boughezal:2016zws,DelDuca:2017twk,
Balitsky:2017flc,Moult:2017jsg,Goerke:2017lei,Balitsky:2017gis,Beneke:2017vpq,Beneke:2017ztn,Feige:2017zci,Moult:2017rpl,
Chang:2017atu,Alte:2018nbn,Beneke:2018gvs,Beneke:2018rbh,Moult:2018jjd,Ebert:2018lzn,Ebert:2018gsn,Bhattacharya:2018vph,Boughezal:2018mvf,vanBeekveld:2019prq,vanBeekveld:2019cks,Bahjat-Abbas:2019fqa,Beneke:2019kgv,Boughezal:2019ggi,Moult:2019mog,Beneke:2019slt}, and ultimately formalize a systematic power expansion of amplitudes and cross sections about these limits.

Recently, using soft collinear effective theory (SCET) \cite{Bauer:2000ew, Bauer:2000yr, Bauer:2001ct, Bauer:2001yt}, subleading power infrared logarithms have been resummed to all orders both for the thrust event shape observable \cite{Moult:2018jjd}, which involves both soft and collinear real radiation, and for the case of threshold logarithms \cite{Beneke:2018gvs,Beneke:2019mua}, which involve only soft real radiation (threshold logarithms were also resummed using diagramatic arguments in \cite{Bahjat-Abbas:2019fqa}).
 However, both these cases considered only subleading power corrections to gluon emission, which, as shown in \cite{Moult:2018jjd}, are related by symmetries to the leading power gluon emission, and therefore exhibit an exponentiation into a standard Sudakov exponential \cite{Sudakov:1954sw}. To understand more generally the structure of subleading power soft and collinear limits, it is necessary to understand cases that go beyond these corrections to soft gluon emission. The most interesting such case is the emission of soft quarks, which generically contribute at leading logarithm at subleading power. Soft quark emission has been used to calculate subleading power corrections in fixed order perturbation theory in Refs.~\cite{Moult:2016fqy,Boughezal:2016zws,Moult:2017jsg,Ebert:2018lzn,Ebert:2018gsn,Boughezal:2018mvf,vanBeekveld:2019cks,Boughezal:2019ggi}, but its all orders structure has not been considered.

In this paper, we initiate a study of the all loop structure of subleading power contributions in the soft and collinear limits that are related to the emission of soft quarks (for example soft quark emission is related to the collinear limit of quark pairs). Through explicit fixed order calculations, we illustrate that the loop level structure of these limits is more non-trivial than for subleading power soft gluon emission, and in particular, that the naive Sudakov exponentiation is violated, even at leading logarithm, by terms proportional to the difference of color generators, $C_A-C_F$. By comparing the results of our fixed order calculation with the organization of the effective theory, we provide insight into the physical nature of the violation of naive Sudakov exponentiation. Interestingly, we show that these terms arise as endpoint divergences in convolution integrals. 

To understand the all loop structure of soft quark emission in a simplified context, we consider $\cN=1$ SUSY QCD (without matter multiplets%
\footnote{Which can be obtained from QCD by setting $C_F\to C_A$ and $n_f \to C_A$.}), 
where fermions are present but these endpoint divergences are absent. 
Here we perform an operator level factorization and renormalization extending the treatment of subleading power resummation in pure gluo-dynamics presented in \cite{Moult:2018jjd}. 
This involves the introduction of several new subleading power jet and soft functions involving soft quarks. 
Using this formalism, we are able to prove that in $\cN=1$ QCD, subleading power logarithms associated with soft quarks exponentiate into a standard Sudakov.%
\footnote{In the context of the fourth order spitting function calculation at $x\to1$ in \cite{Soar:2009yh} it was observed that the $\cN=1$ SUSY limit of QCD has a different structure of leading logarithmic terms at next to leading power in the $x\to1$ expansion, revealing a non trivial contribution of supersymmetry breaking terms at leading log. This observation is consistent with our connection of supersymmetry breaking terms to endpoint divergences, since such divergences will appear in the $x\to 1$ limit at subleading power.}

Although we do not here solve the problem of endpoint divergences in QCD, based on our fixed order calculations, combined with intuition from the case of $\cN=1$ QCD, we conjecture a form for the leading logarithmic ``soft quark Sudakov" in QCD. Our result takes a simple, but interesting form with similarities to structures observed in the large-x resummation of off-diagonal splitting functions  \cite{Soar:2009yh,Vogt:2010cv,Almasy:2010wn,Presti:2014lqa,Almasy:2015dyv}.  We also show that this soft quark Sudakov seems to have some degree of universality. Our paper therefore provides three new ingredients towards understanding factorization at subleading power
\begin{itemize}
\item An explicit calculation of the subleading power collinear limits involving quarks for $e^+e^-\to 3$ parton and $H\to 3$ parton amplitudes at two loops.
\item An operator level factorization and resummation of subleading power logarithms in $\cN=1$ QCD, involving new jet and soft functions with soft quarks that go beyond those considered in \cite{Moult:2018jjd}.
\item A conjecture for the full leading logarithmic soft quark Sudakov in QCD.
\end{itemize}
These provide a considerable step beyond the resummation of subleading power corrections to soft gluon emission that have been considered to date. In particular, we hope that by highlighting that endpoint divergences can have an effect even at leading logarithmic order, and  by giving them a physical interpretation as arising when there are fields in different color representations, we provide some insight into an eventual complete field theory solution of subleading power resummation in QCD.

An outline of this paper is as follows.  In \Sec{sec:nlp_collinear} we study the subleading power collinear limits of amplitudes at loop level, highlighting interesting effects that arise when considering collinear limits that do not exist at leading power. We also discuss the relation of these results to the violation of the most naive anticipated form of factorization at subleading power (Sudakov exponentiation). In \Sec{sec:n1} we perform an operator level factorization and resummation in $\cN=1$ QCD, building on our work in \cite{Moult:2018jjd}. In \Sec{sec:soft_quark_sudakov} we conjecture a form for the leading logarithmic soft quark Sudakov in QCD. We conclude  in \Sec{sec:conc}, and discuss some remaining open issues for achieving a complete understanding of leading logarithmic subleading power resummation in QCD.

\section{Subleading Power Collinear Limits at Loop Level}\label{sec:nlp_collinear}

In this section we explicitly compute the subleading power collinear limits of $e^+e^-\to 3$ partons and $H\to 3$ partons and study the behavior of the limits that do not have a leading power analogue. The matrix elements for $e^+e^-\to 3$ partons and $H\to 3$ partons were computed at two loops in \cite{Garland:2001tf,Garland:2002ak,Gehrmann:2011aa}, and we obtain our results by expanding these amplitudes in the collinear limits. Details of how this expansion is performed were given in \cite{Moult:2018jjd}. The results of \cite{Garland:2001tf,Garland:2002ak,Gehrmann:2011aa} provide the complete color structure, so that in addition to QCD, we can obtain the result in $\cN=1$ SUSY with an adjoint gluino by setting $C_F \to C_A$ and $n_f \to C_A$.

Note that there has recently been significant interest in Higgs form factors from the perspective of maximal transcendentality, and their relation to similar form factors in $\cN=4$ \cite{Brandhuber:2017bkg,Brandhuber:2018xzk,Brandhuber:2018kqb,Jin:2018fak,Jin:2019ile}, which have primarily focused on the $C_F\to C_A$ limit. Here we are interested in the structure of the maximally transcendental terms that persist in the $C_F\to C_A$ limit.

\subsection{Physical Observables from Consistency Relations}\label{sec:consistency}

We are ultimately interested in the behavior of subleading power corrections to infrared event shape observables. Such event shapes are complicated to directly compute perturbatively, since they involve both real and virtual corrections. An efficient way to obtain the leading logarithms is to use consistency relations derived in \cite{Moult:2016fqy,Moult:2017jsg}, which allow the leading logarithms to be expressed entirely in terms of virtual corrections that can computed efficiently using known amplitudes. Here we briefly review this approach to make clear why we focus on a particular set of virtual corrections. More details can be found in \cite{Moult:2016fqy,Moult:2017jsg}.

The $N$-loop fixed order result for an event shape $\tau$ at next-to-leading power (NLP), i.e. terms that scale like $\tau^{0}$ modulo logarithms, can be written as \cite{Moult:2016fqy,Moult:2017jsg}
\begin{align}\label{eq:constraint_setup}
\frac{1}{\sigma_0}\frac{\df\sigma^{(2,N)}}{\df\tau}
  = & \sum_{\kappa}\sum_{i=0}^{2N-1} \frac{c_{\kappa,i}}{\epsilon^i} \left( \frac{\mu^{2N}}{Q^{2N} \tau^{m(\kappa)}}   \right)^\epsilon
 + \dots
\,.\end{align}
The subleading poles in $\epsilon$ are only relevant beyond LL order, and will not be considered here. The sum in \Eq{eq:constraint_setup} is over $\kappa$, which denotes different combinations of soft, collinear, or hard loops (we will often use the word ``loops'' generically, such that it includes both loop and phase space integrals). The power of $\tau$ appearing in the result, denoted $m(\kappa)$ is determined by the number of soft, collinear or hard loops. As a concrete example, at one-loop we can have either
\begin{align} \label{eq:classes1}
\text{soft:} \qquad &\kappa=s\,, \qquad m(\kappa) =2\,, \\
\text{collinear:} \qquad &\kappa=c\,, \qquad m(\kappa)=1 \,.\nn
\end{align}
For an infrared safe observable, the constraint that all infrared poles in $\epsilon$ cancel places strong constraints on the coefficients of different poles. It can be shown (see \cite{Moult:2016fqy,Moult:2017jsg} for details) that the leading logarithmic result can be expressed entirely in terms of the hard-collinear contributions, i.e. those with one collinear loop, and $N-1$ hard loops. 
This occurs because these contributions have a unique scaling with the observable, of $\tau^{-\epsilon}$ in \eq{constraint_setup}.
Explicitly, 
\begin{align}\label{eq:constraints_final}
\frac{1}{\sigma_0}\frac{\df\sigma^{(2,N)}}{\df\tau}
&= c_{hc,2N-1} \log^{2N-1} \tau +\cdots
\,.\end{align}
Here  $c_{hc,2N-1}$ denotes the leading pole in $\epsilon$ with $N-1$ hard loops correcting a single collinear splitting. Because of this, we can gain insight into the structure of the leading logarithms for infrared safe observables by studying the virtual corrections to the subleading power collinear limits of squared amplitudes. These must then be integrated over the simple two particle collinear phase space to obtain the result. This will enable us, as shown in \app{PSintegral}, to compute subleading power corrections to event shape observables to $\cO(\alpha_s^3)$ analytically, which is not feasible directly. Additionally, the study of these virtual corrections is interesting in its own right.

\subsection{$e^+e^-\to 3$ Partons}\label{sec:ee}

We begin by looking at the case of $e^+e^-\to q \bar q g$. Here we will be interested in the limit where the two quarks become collinear, which does not have a leading power analog. The analog of the analysis below where the gluon becomes collinear to a quark, has been considered in \cite{Moult:2018jjd}. We label the kinematics as
\begin{align}\label{eq:fig_ee}
\fd{12cm}{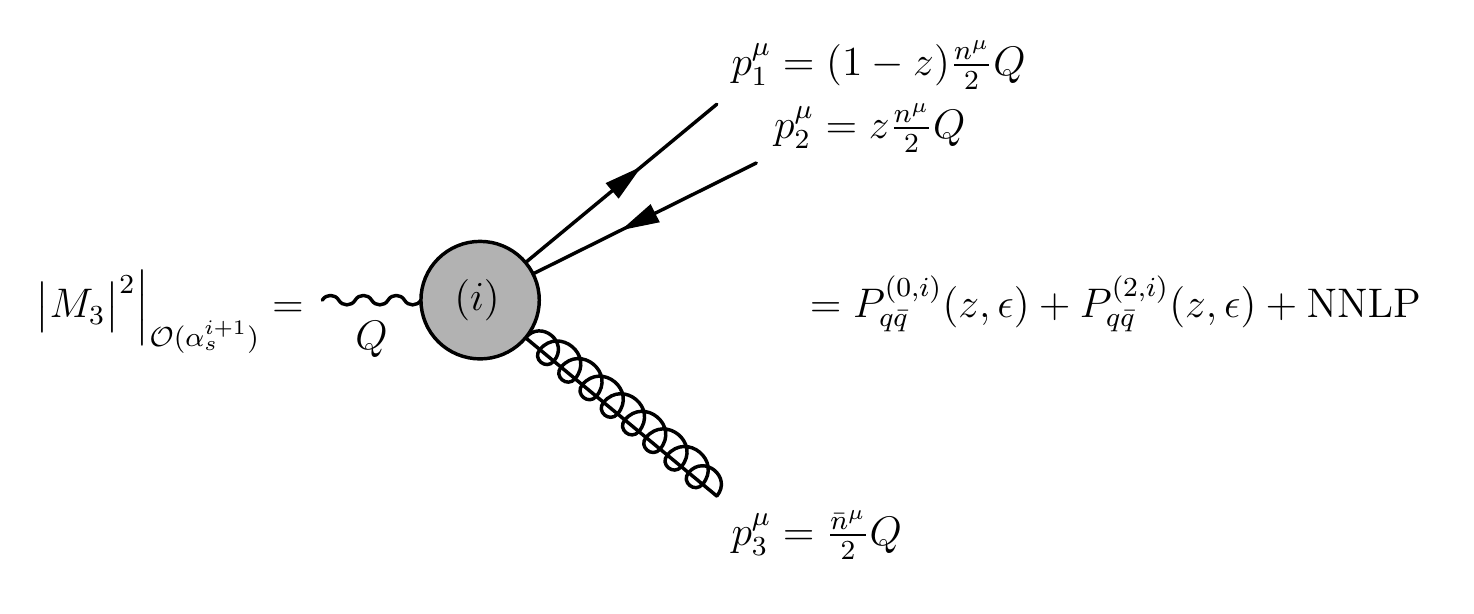}\,,
\end{align}
and the superscripts in $P_{q\bar q}^{(i,j)}$ are the power $i$ in the collinear expansion, and the order $j$ in the $\alpha_s$ expansion and here we are always keeping the number of external particles fixed, hence higher orders in $\alpha_s$ indicate higher loop corrections.
Note that we have
\begin{align}
	P_{q\bar q}^{(0,i)} (z,\epsilon)= 0 \qquad\forall \,i\,,
\end{align}
meaning that the leading power (LP) contribution to the limit where the two quarks become collinear vanishes at all orders in $\alpha_s$.
Hence, as anticipated, this limit starts at next-to-leading-power (NLP).
At NLP at tree level, we write the result as
\begin{align}\label{eq:Pqq20}
	P_{q\bar q}^{(2,0)} (z,\epsilon)= 8N_c g_s^2 C_F  \left( \frac{1-z}{z}+\frac{z}{1-z}   \right).
\end{align}
Here the exponent ${(2,0)}$ labels that this is subleading power, and tree level.

We now consider loop corrections to this result. We keep only the highest transcendental terms, which are obtained by assigning $1/\epsilon^n$, $\log(z)^n$ and $\log(1-z)^n$ transcendental weight $n$ and keeping only the weight 2 part of the result, and ignore weight 2 functions which are not logarithmic divergent in the $z\to 0$ or $z\to1$ limit, such as $\Li_2(z)$, since these will not contribute leading logarithms once integrated against the collinear phase space. At one-loop, we find
\begin{align} \label{eq:Pqq21}
P_{q\bar q}^{(2,1)}(z,\epsilon)=&P_{q\bar q}^{(2,0)}(z,\epsilon) \frac{\alpha_s}{4\pi} \mu^{2\epsilon} \Big[ 4 {\bf{T}}_1 \cdot {\bf{T}}_3    \frac{ [(1-z)Q^2]^{-\epsilon}}{\epsilon^2} +4 {\bf{T}}_1 \cdot {\bf{T}}_2 \frac{s^{-\epsilon}}{\epsilon^2}  +4 {\bf{T}}_2 \cdot {\bf{T}}_3 \frac{[zQ^2]^{-\epsilon}}{\epsilon^2} \nn \\
&  +4 {\bf{T}}_1 \cdot {\bf{T}}_2 \left(  \frac{[Q^2]^{-\epsilon}}{\epsilon^2}-  \frac{[z(1-z)Q^2]^{-\epsilon}}{\epsilon^2}  \right) -4 {\bf{T}}_1 \cdot {\bf{T}}_2  \left(  \frac{s^{-\epsilon}}{\epsilon^2}-  \frac{[z(1-z)s]^{-\epsilon}}{\epsilon^2}  \right)  \Big]\nn \\
&+\cO\left(\frac{1}{\epsilon}\right)\,,
\end{align}
where $s=(p_1+p_2)^2$ is the invariant mass of the $q\bar{q}$ pair. Note that since $p_1^\mu$ and $p_2^\mu$ are collinear to the same direction, their invariant mass $s$ is much smaller than the hard scale of the process, which we identify with $Q$. In formulae
\be
	s=(p_1 + p_2)^2 \sim Q^2\tau \ll Q^2 = (p_1 + p_2 + p_3)^2 \,,
\ee
where $\tau$ is an infrared power counting parameter that we will ultimately identify with a physical observable.
In terms of Casimirs we have the following expressions for the color generators
\begin{align}\label{eq:Casimirsee}
{\bf{T}}_1 \cdot {\bf{T}}_2=\frac{C_A}{2}-C_F \,, \qquad {\bf{T}}_1 \cdot {\bf{T}}_3=-\frac{C_A}{2}  \,, \qquad {\bf{T}}_2 \cdot {\bf{T}}_3=-\frac{C_A}{2}\,.
\end{align}
Note that, even though each term in \eq{Pqq21} carries a $1/\epsilon^2$, only the terms in the first line of  \eq{Pqq21} are predicted by Catani's dipole formula \cite{Catani:1998bh} and its generalization \cite{Dixon:2008gr,Becher:2009qa,Gardi:2009qi,Almelid:2015jia}. 
The other terms in \eq{Pqq21}, after expanding in $\epsilon$, will generate $1/\epsilon^2$ and $1/\epsilon$ contributions that cancel against each other, leaving only finite terms that would belong to the remainder part of the amplitude. However, they contribute at leading logarithmic order once integrated over the collinear phase space.
The fact that this can happen is generic and not a new finding of this paper. A more detailed discussion of this point can be found, for example, in Sec. 5 of \cite{Moult:2018jjd}.

When integrating \eq{Pqq21} over collinear phase space, the leading result will come from the  $z\to 0$, or $z\to 1$ limits. It is therefore interesting to study the behavior of this result in these limits. Since the result is symmetric in $z\leftrightarrow 1-z$, we consider $z\to 0$. We find
\begin{align}\label{eq:qq_collinear}
	\left.P_{q\bar q}^{(2,1)}\right|_{z\to 0} &= \\
	&\hspace{-1cm} \left.P_{q\bar q}^{(2,0)}\right|_{z\to 0} \frac{\alpha_s}{4\pi} \mu^{2\epsilon}  \Big[  4({\bf{T}}_1+{\bf{T}}_2) \cdot {\bf{T}}_3   \frac{[Q^2]^{-\epsilon}}{\epsilon^2}  +4({\bf{T}}_1-{\bf{T}}_3) \cdot {\bf{T}}_2 \left(  \frac{[Q^2]^{-\epsilon}}{\epsilon^2}-  \frac{[zQ^2]^{-\epsilon}}{\epsilon^2}  \right)  \Big ]\,\nn\\
	&\hspace{-1.72cm}=  32\pi N_c C_F \left(\frac{\alpha_s}{4\pi}\right)^2 \left(\frac{\mu^2}{Q^2}\right)^{\epsilon} \left( \frac{1}{z}\right) \Big[  4({\bf{T}}_1+{\bf{T}}_2) \cdot {\bf{T}}_3   \frac{1}{\epsilon^2}  +4({\bf{T}}_1-{\bf{T}}_3) \cdot {\bf{T}}_2 \left(  \frac{1}{\epsilon^2}-  \frac{[z]^{-\epsilon}}{\epsilon^2}  \right)  \Big ]\,. \nn
\end{align}
This result is quite interesting. The naive intuition from leading power factorization is that the leading double logs (equivalently the leading poles in $\epsilon$) arise from soft gluons that view the two collinear quarks coherently, and are insensitive to their energy fractions. This would lead to a cusp like result \cite{Korchemsky:1987wg,Korchemskaya:1992je} with color structure $({\bf{T}}_1+{\bf{T}}_2) \cdot {\bf{T}}_3 $ (in this case an adjoint cusp), as illustrated schematically by
\begin{align}\label{eq:color_coherence}
\fd{3cm}{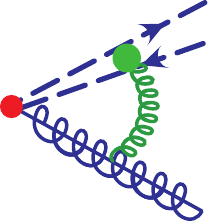}\,.
\end{align}
If $P_{q\bar q}^{(2,0)}(z,\epsilon)$ were non-singular as $z\to 0$, then the second term in \Eq{eq:qq_collinear} could be expanded in $\epsilon$, and would not contribute to the leading logarithm, and we would have exactly this picture. However, instead, we get a contribution from $z\to 0$ that violates this naive expectation. We will refer to this as an endpoint contribution. Note that if $ {\bf{T}}_1={\bf{T}}_3$, as would occur in $\cN=1$ SUSY QCD, then the naive factorization picture holds. We will discuss the interpretation in terms of factorization in SCET in \Sec{sec:factorization}.

This analysis can be extended to two-loops\footnote{Note that $P_{q\bar q}^{(2,n)}$ is the NLP $\cO(\alpha_s^{n+1})$ contribution to the amplitude squared in the collinear limit. Therefore, $P_{q\bar q}^{(2,2)}$ contains both the subleading two loop correction interfering with the tree level as well as the squared one loop subleading splitting function. Note also that since next-to-leading power corrections start with one real emission,  $P_{q\bar q}^{(i,2)}$ is part of the $\cO(\alpha_s^3)$ RVV terms contributing to color singlet decay.}. Remarkably, we find that the one loop result simply exponentiates
\begin{align}\label{eq:p2qqbar}
P_{q\bar q}^{(2,2)}(z,\epsilon)=&P_{q\bar q}^{(2,0)}(z,\epsilon) \frac{1}{2!} \left( \frac{\alpha_s}{4\pi} \right)^2 \mu^{4\epsilon} \Big[ 4 {\bf{T}}_1 \cdot {\bf{T}}_3    \frac{ [(1-z)Q^2]^{-\epsilon}}{\epsilon^2} +4 {\bf{T}}_1 \cdot {\bf{T}}_2 \frac{s^{-\epsilon}}{\epsilon^2}  +4 {\bf{T}}_2 \cdot {\bf{T}}_3 \frac{[zQ^2]^{-\epsilon}}{\epsilon^2} \nn \\
&\hspace{-1.5cm}  +4 {\bf{T}}_1 \cdot {\bf{T}}_2 \left(  \frac{[Q^2]^{-\epsilon}}{\epsilon^2}-  \frac{[z(1-z)Q^2]^{-\epsilon}}{\epsilon^2}  \right) -4 {\bf{T}}_1 \cdot {\bf{T}}_2  \left(  \frac{s^{-\epsilon}}{\epsilon^2}-  \frac{[z(1-z)s]^{-\epsilon}}{\epsilon^2}  \right)  \Big]^2\,.
\end{align}
The $\cO(\epsilon^0 \ln^2)$ of this result can be checked against the expansion of the analytic amplitudes (squared) for $e^+e^- \to 3$ jets at NNLO \cite{Garland:2001tf}. Given the size of these $\cO(\alpha_s^3)$ expressions, the extraction of the subleading collinear limit from them is non trivial. However, to extract only the leading transcendental part is quite simple. The matrix elements squared is written in terms of rational functions multiplying transcendental functions. We can simplify the procedure of taking the collinear limit of such expressions by performing the expansion of the transcendental functions at the symbol level.\footnote{Note that taking the symbol of an expression loses certain information, such as the zeta value terms. However, such terms are less singular in the end point and do not contribute to the LL considered here.} Then the only functions that are needed beyond the leading collinear limit are the rational functions, whose expansion is straightforward. Alternatively, the result of \cite{Garland:2001tf} is expressed in terms of two-dimensional harmonic polylogarithms that can be directly expanded using \cite{Panzer:2014caa}.

Coming back to the result of \eq{p2qqbar}, we see from explicit calculation that at least to two loops, loop corrections to the collinear limit behave as expected from naive factorization, with the exception of endpoint contributions.
These endpoint contributions are proportional to $(C_A-C_F)^n$ and therefore vanish in $\cN=1$ SUSY. 
In \Sec{sec:n1}, we will perform an operator level factorization analysis to show that in $\cN=1$ QCD, the leading logarithms exponentiate to all orders. 
In QCD, we are not yet able to perform such an analysis due to these endpoint contributions. 
Nevertheless, we give a conjecture for the form of the soft quark Sudakov in \Sec{sec:soft_quark_sudakov}.

\subsection{$H\to 3$ Partons}\label{sec:H}

We can also consider the $\lambda HG^{\mu\nu}G_{\mu\nu}$ effective operator producing
the decay of $H\to q\bar q g$, in the limit where the quark and gluon become collinear, which much like the collinear limit of two quarks in $e^+e^-\to q \bar q g$ does not have a leading power limit. Here we will find an identical story to that for the $q\bar q$ collinear limit just considered, providing some evidence that it is generic.

We label the partons for the Higgs decay as
\begin{align}
\fd{12cm}{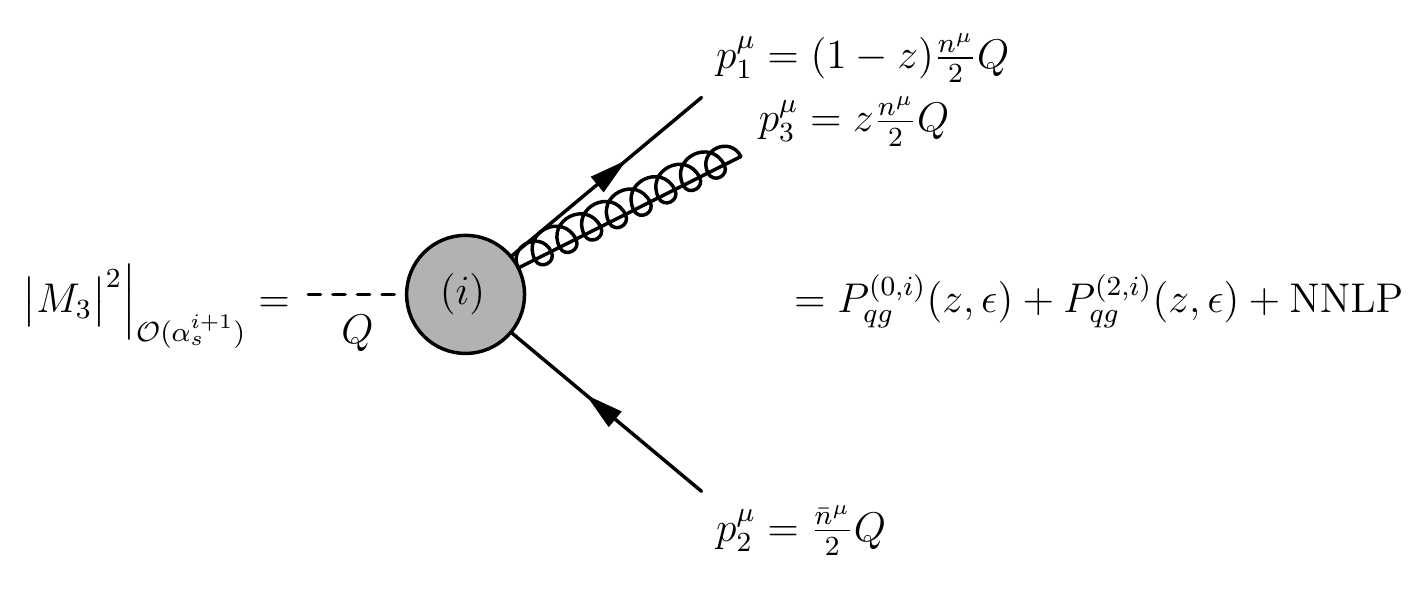}\,.
\end{align}
The LO NLP splitting function is
\begin{align}
P_{qg}^{(2,0)}(z,\epsilon)=32 \pi^2 \lambda^2 Q^2 C_F \frac{z^2}{1-z}\,.
\end{align}
At one loop, we find
\begin{align}
P_{qg}^{(2,1)}(z,\epsilon)&=P_{qg}^{(2,0)}(z,\epsilon) \frac{\alpha_s}{4\pi} \mu^{2\epsilon} \Big[ 4 {\bf{T}}_1 \cdot {\bf{T}}_2    \frac{ [(1-z)Q^2]^{-\epsilon}}{\epsilon^2} +4 {\bf{T}}_1 \cdot {\bf{T}}_3 \frac{s^{-\epsilon}}{\epsilon^2}  +4 {\bf{T}}_2 \cdot {\bf{T}}_3 \frac{[zQ^2]^{-\epsilon}}{\epsilon^2} \nn \\
&\hspace{-1cm}  +4 {\bf{T}}_1 \cdot {\bf{T}}_3 \left(  \frac{[Q^2]^{-\epsilon}}{\epsilon^2}-  \frac{[z(1-z)Q^2]^{-\epsilon}}{\epsilon^2}  \right) -4 {\bf{T}}_1 \cdot {\bf{T}}_3  \left(  \frac{s^{-\epsilon}}{\epsilon^2}-  \frac{[z(1-z)s]^{-\epsilon}}{\epsilon^2}  \right) \Big]\,.
\end{align}
In terms of Casimirs we have
\begin{align}
{\bf{T}}_1 \cdot {\bf{T}}_2=\frac{C_A}{2}-C_F \,, \qquad {\bf{T}}_1 \cdot {\bf{T}}_3=-\frac{C_A}{2}  \,, \qquad {\bf{T}}_2 \cdot {\bf{T}}_3=-\frac{C_A}{2}\,.
\end{align}
We can now consider the behavior in the $z\to 1$ limit, which gives rise to the leading logarithm when integrated over $z$. In the $z\to 1$ limit, we have
\begin{align}\label{eq:qg_collinear}
\left.P_{qg}^{(2,1)}\right|_{z\to 1}&=P_{qg}^{(2,0)} \frac{\alpha_s}{4\pi} \mu^{2\epsilon} \Big[ 4({\bf{T}}_1+{\bf{T}}_3) \cdot {\bf{T}}_2   \frac{[Q^2]^{-\epsilon}}{\epsilon^2}  +4({\bf{T}}_3-{\bf{T}}_2) \cdot {\bf{T}}_1 \left(  \frac{[Q^2]^{-\epsilon}}{\epsilon^2}-  \frac{[(1-z)Q^2]^{-\epsilon}}{\epsilon^2}  \right) \Big]\,.
\end{align}
Here we again see an identical structure as found in the case of $e^+e^-$ decay. Away from $z\to 1$, we can expand the second term in \Eq{eq:qg_collinear}, and find the expected result  from a cusp with color structure $({\bf{T}}_1+{\bf{T}}_3) \cdot {\bf{T}}_2$. However, there is another contribution from the endpoint $z\to 1$.
Again, if $ {\bf{T}}_3={\bf{T}}_2$, as would occur in $\cN=1$ QCD, then the second term vanishes and naive factorization holds.

This result can also be extended to two loops. As with the case of the $e^+e^-$ decay, we find that it exponentiates
\begin{align}\label{eq:qg_collinear_2}
P_{qg}^{(2,2)}&=P_{qg}^{(2,0)} \frac{1}{2!} \left( \frac{\alpha_s}{4\pi}\right)^2 \mu^{4\epsilon} \Big[ 4 {\bf{T}}_1 \cdot {\bf{T}}_2    \frac{ [(1-z)Q^2]^{-\epsilon}}{\epsilon^2} +4 {\bf{T}}_1 \cdot {\bf{T}}_3 \frac{s^{-\epsilon}}{\epsilon^2}  +4 {\bf{T}}_2 \cdot {\bf{T}}_3 \frac{[zQ^2]^{-\epsilon}}{\epsilon^2} \nn \\
&  +4 {\bf{T}}_1 \cdot {\bf{T}}_3 \left(  \frac{[Q^2]^{-\epsilon}}{\epsilon^2}-  \frac{[z(1-z)Q^2]^{-\epsilon}}{\epsilon^2}  \right) -4 {\bf{T}}_1 \cdot {\bf{T}}_3  \left(  \frac{s^{-\epsilon}}{\epsilon^2}-  \frac{[z(1-z)s]^{-\epsilon}}{\epsilon^2}  \right) \Big]^2\,.
\end{align}
We conjecture that it exponentiates to all loops, however, we are only able to prove this in $\cN=1$ QCD. 
Note that the $\cO(\epsilon^0)$ expansion of the result in \eq{qg_collinear_2} can be cross checked against the subleading power collinear expansion of the matrix elements squared constructed from the analytic result of the amplitudes for $H\to 3$ partons at NNLO \cite{Gehrmann:2011aa}.
The procedure is analogous as the one we explained for the case of $e^+ e^- \to 3$ jets below \eq{p2qqbar}.

\subsection{Insights into Factorization and Endpoint Divergences}\label{sec:factorization}

In this section we briefly discuss the interpretation of the above fixed order calculation in terms of subleading power factorization formulas in SCET. This section assumes significant familiarity with SCET at subleading power, which we do not attempt to review in this paper. Details building up the formalism for subleading power factorization used here can be found in \cite{Kolodrubetz:2016uim,Feige:2017zci,Moult:2017rpl,Moult:2018jjd,Moult:2019mog}.

Factorization formulas at subleading power generically involve convolutions in the label momentum of collinear partons, or in certain momenta of soft partons. A general issue is that these convolutions (at least when treated naively) can exhibit divergences. It is currently not understood when such divergences occur, and when they do occur how to overcome them to achieve subleading power resummation. While such divergent convolution integrals are expected to occur somewhat generically at next-to-leading logarithm, we believe that focusing on cases where they occur at leading logarithm may be useful for understanding their physical origin. 

For the case of soft gluon emission at subleading power for dijet event shapes, it was shown that such divergences do not occur at leading logarithm due to reparametrization invariance. In the case of $\cN=1$ SUSY, they also do not occur for soft quark emissions due to supersymmetry. This constrains such divergences to be proportional to $(C_A-C_F)^n$ at leading logarithm in QCD. In this section, we briefly discuss the physical picture of such divergences, which we hope will lead to an understanding in the near future. In addition to this insight into the structure of the endpoint contributions, we also wish to emphasize that it is generally assumed that leading logarithms can be derived by considering anomalous dimensions that are diagonal in the convolution variables. This provides an example where this does not occur in this naive manner.

To understand the structure of the endpoint divergences appearing in our fixed order calculations, we consider how these limits are factorized in the EFT. The matching for the relevant operators was studied in detail for Higgs decay in \cite{Moult:2017rpl} and for $e^+e^-$ annihilation in  \cite{Feige:2017zci}. We consider first the case of Higgs decay, where a quark and gluon become collinear. The full theory diagram that contributes to the matching, expanded to $\cO(\lambda)$ in the power counting is given by
\begin{align}\label{eq:Higgs_match}
\left. \fd{2.5cm}{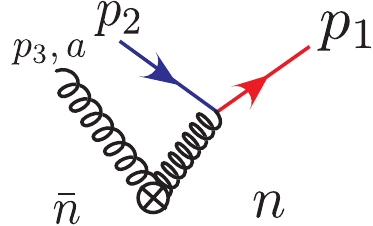}\right |_{\cO(\lambda)} &= \frac{-2ig\omega_3}{\omega_2}  \bar u_n (p_1) \Sl{\epsilon}_{3\perp} T^a v_{\bar n}(p_2)\,.
\end{align}
Here $\omega_1 \equiv \bn \cdot p_{1}$ is the large component of the $n$-collinear momentum $p_1^\mu$ and $\omega_{2,3} \equiv n \cdot p_{2,3}$ are the large component of the $\bn$-collinear momenta $p_2^\mu$ and $p_3^\mu$. The large component of the momentum is usually referred as the \emph{label} momentum of the particle.
This is matched in the EFT onto an operator
\begin{align}\label{eq:Higgs_gq_operator}
\cO^{(1)}_{\cB \bar n}= -2g\, \frac{\omega_3}{\omega_2}\,
  \bar \chi_{n,\omega_1} \Sl{\cB}_{\perp \bar n, \omega_3} 
    \chi_{\bar n,-\omega_2} H\,,
\end{align}
whose Feynman rule is given by
\begin{align}
\fd{3cm}{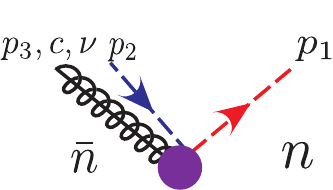}
  &=-2ig T^c \frac{\omega_3 }{\omega_2}   \left( \gamma^\nu_\perp-\frac{\Sl{p}_{3\perp} n^\nu}{\omega_3} \right)\,.
\end{align}
The matching coefficient for this operator exhibits a divergence as $\omega_2\to 0$. In this limit, the off-shell gluon in the diagram in \Eq{eq:Higgs_match} that is being integrated out becomes on-shell, and should no longer be incorporated in the hard matching coefficient, but rather in the long distance emission of the soft quark. In the effective theory, this overlap should be removed by the zero bin \cite{Manohar:2006nz}, however, the exact implementation of the zero bin in this case is not fully understood.

This complication generally does not arise at leading power.
At leading power both for Higgs decay and $e^+e^-$ annihilation, we have only one field per collinear direction in the hard scattering operator and its label momentum, let's call it $\omega$, is fixed by momentum conservation to be a hard scale of the process, $\omega = Q$. 
The reason why here it is meaningful to talk about the singular behavior of the Wilson coefficient as a label momentum goes to zero is because at subleading power we can have multiple collinear fields in one collinear sector (as in the operator in \eq{Higgs_gq_operator} where both $\Sl{\cB}_{\perp \bar n, \omega_3}$ and $\chi_{\bar n,-\omega_2}$ belong to the same $\bn$-collinear sector and have label momenta $\omega_3$ and $\omega_2$ respectively).
Therefore, momentum conservation for the label components constrains only the sum of the labels in a collinear sector, hence in our example we have $\omega_2 + \omega_3 = Q $. 
The relative size of $\omega_2$ and $\omega_3$ can be arbitrary and it is necessary to integrate over all possible configurations, including the one where one label momentum vanishes as the other becomes closer and closer to the hard scale.

This presence of a singularity in the matching coefficient is completely generic, and occurs also for $e^+e^-$ annihilation. For the case that the two quarks become collinear, we have
\begin{align}\label{eq:matching_qqsame1}
\left( \fd{3cm}{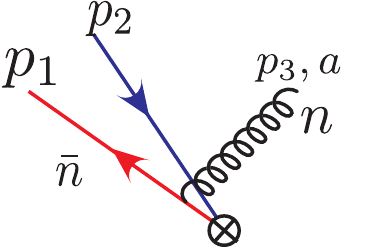} \right)  \left. \vphantom{\fd{3cm}{figures/matching_subleadingvertex_casec1_low}}   \right|_{\mathcal{O}(\lambda)}
 &= - \frac{g}{\omega_1}\bar u_{\bar n}(1) T^a \Sl{\epsilon}^*_\perp \frac{\Sl{n}}{2}  \Gamma     v_{\bar n}(2)\,,
 \\
\left( \fd{3cm}{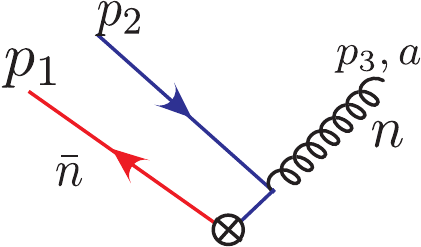} \right)  \left. \vphantom{\fd{3cm}{figures/matching_subleadingvertex_casec2_low}}   \right|_{\mathcal{O}(\lambda)}
 &= 
 + \frac{g}{\omega_2 }  \bar u_{\bar n}(1) T^a \Gamma  \frac{\Sl{n}}{2} \Sl{\epsilon}^*_\perp  v_{\bar n}(2)
 \,. \nn
 \end{align}

To resum subleading power logarithms, one must renormalize these operators involving multiple collinear fields in each collinear sector. It is generically assumed that at leading logarithm, this renormalization is diagonal in the momentum fractions of the collinear fields, since it should arise from diagonal radiation which sees the two collinear fields coherently, as was illustrated in \Eq{eq:color_coherence}.
Looking at our results from the fixed order calculation in this light (and recalling that $z=\omega_1 Q$, $(1-z)=\omega_2 Q$), 
\begin{align}
	\left.P_{q\bar q}^{(2,1)}\right|_{z\to 0}=&P_{q\bar q}^{(2,0)} \frac{\alpha_s}{4\pi} \left(\frac{\mu^2}{Q^2}\right)^{\epsilon}  \Big[  4({\bf{T}}_1+{\bf{T}}_2) \cdot {\bf{T}}_3   \frac{1}{\epsilon^2}  +4({\bf{T}}_1-{\bf{T}}_3) \cdot {\bf{T}}_2 \left(  \frac{1}{\epsilon^2}-  \frac{z^{-\epsilon}}{\epsilon^2}  \right) \Big ]\nn\\
	+& \cO\left(\frac{1}{\epsilon}\right) \,.
\end{align}
The first term in square brackets is proportional to the sum of the color charges of the fields in the collinear sector, as expected for the usual diagonal terms. However the second term is not, and due to the $1/z$ behavior of $P_{q\bar q}^{(2,0)}$ it contributes a leading logarithmic contribution from $z\to 0$.

To understand physically what is occurring, it is instructive to look at the structure of the cusp at the hard scattering vertex for $\omega_2>0$ and as $\omega_2\to 0$. For $\omega_2>0$, in the Higgs case, we have a fundamental cusp
\begin{align}
\fd{3cm}{figures/matching_lam_qqg_Feyn_low}\to \fd{3cm}{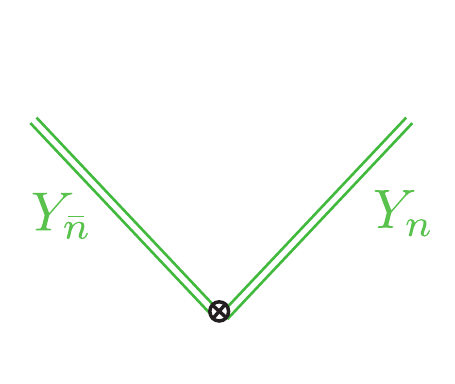}\,,
\end{align}
whereas, as $\omega_2\to 0$, for the long distance contribution, we get an adjoint cusp
\begin{align}
\fd{3cm}{figures/matching_lam_qqg_low}\to \fd{3cm}{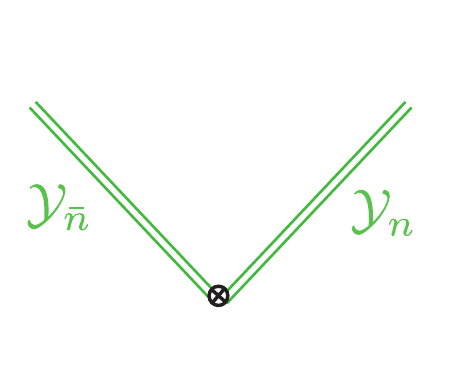}\,.
\end{align}
Therefore, there is a non-smooth transition in the color representation of the cusp as we approach the limit $\omega_2 \to 0$. 
Note that if the quarks are in the adjoint, as occurs in $\cN=1$ SUSY, one has a fundamental cusp in both cases, and so the limit is smooth.

In the cases where factorization works naively, namely for subleading corrections to soft gluon emission, or soft quark emission in $\cN=1$ SUSY, one does not have this issue of a non-smooth limit for the representation of the cusp. We believe that this clearly demonstrates that this issue is related to the treatment of the zero bin, which apparently must have a non-trivial structure related to the changing of the color representations of the cusp.  An interesting upshot of our analysis is that we have taken a step towards better understanding the nature of endpoint singularities by showing that at LL, they must be proportional to powers of $C_A-C_F$. This physical picture also strongly suggests that the result is still related at LL to the cusp anomalous dimension, albeit perhaps in different representations.  While we do not at this stage have a complete understanding of the treatment of the zero-bin overlap region, we hope that this analysis provides an insight into its structure. We have also highlighted a particularly clear example where it is important even at LL, violating the standard lore.

Although we will restrict our attention in this paper to NLP, the loop level structure of the collinear limits given in \Secs{sec:ee}{sec:H} persists to all powers, with only a modification of the tree level coefficient. This suggests that at LL, endpoint contributions in $\cN=1$ QCD are absent to all powers, and more generally, that a key difference between $\cN=1$ and QCD at LL to all powers is endpoint contributions proportional to $(C_A-C_F)^n$. It is clear that an understanding of the end point contributions will also be a necessary ingredient to carry out resummation at higher orders in the power counting.

\section{Factorization and Resummation in $\cN=1$ SUSY QCD}\label{sec:n1}

In this section, as an example and extension of the formalism developed in \cite{Moult:2018jjd}, we perform an operator level factorization and renormalization in $\cN=1$ QCD to resum subleading power infrared logarithms. Since many of the relevant details have been worked out in detail in \cite{Moult:2018jjd}, here we focus on the structure of the operators and the structure of the renormalization group mixing. This analysis extends beyond the case of pure gluo-dynamics considered in \cite{Moult:2018jjd} through the inclusions of a number of new soft and jet functions involving quarks. We believe that this provides one more step towards a complete understanding of subleading power factorization at leading logarithm. In particular from our explicit calculations, in \Sec{sec:nlp_collinear}, it appears that at leading logarithm the only difference between $\cN=1$ and full QCD is the presence of endpoint divergences. 

The underlying reason for the simplification in $\cN=1$ is that supersymmetry relates quark and gluon emissions. This has been studied in detail in the abelian case in \cite{Dumitrescu:2015fej}, and it would be interesting to understand it in more detail at an operator level in SCET. However, we leave this to future work.

For concreteness, in this section we will consider the event shape thrust \cite{Farhi:1977sg}, although the identical set of operators and renormalization group structure would also apply for any other \SCETi dijet event shape. Here we define the dimensionful thrust observable as $\Tau$, and the dimensionless version as $\tau=\Tau/Q$, where $Q$ is the center of mass energy of the scattering. In the $\tau\to 0$ limit, the thrust cross section can be expanded in powers of $\tau$ as
\begin{align}\label{eq:intro_expansion}
\frac{\df\sigma}{\df\tau} &=\frac{\df\sigma^{(0)}}{\df\tau} +\frac{\df\sigma^{(1)}}{\df\tau} +\frac{\df\sigma^{(2)}}{\df\tau}+\frac{\df\sigma^{(3)}}{\df\tau} +{\cal O}(\tau)\,.
\end{align}
Here the cross section $\df\sigma^{(n)}/\df\tau$ describes all terms in $\alpha_s$ that scale like $\tau^{n/2-1}$. For thrust, $\df\sigma^{(n)}/\df\tau=0$ for $n$ odd. Here we will focus on $n=2$, which is also known as the next-to-leading power (NLP) cross section. The logarithmic structure of the NLP cross section is given by
\begin{align}\label{eq:subl_expansion}
\frac{1}{\sigma_0}\frac{d\sigma^{(2)}}{d\tau} 
  =\sum\limits_{n=1}^\infty \sum\limits_{m=0}^{2n-1} \left(  \frac{\alpha_s(\mu)}{4\pi} \right)^n c^{(2)}_{n,m}\, \,\log^{m}(\tau)\,.
\end{align}
In this section we will extend our analysis in \cite{Moult:2019mog} to derive the structure of the leading logarithms in the NLP expression for $e^+e^-$ annihilation and Higgs decay in $\cN=1$ QCD. In particular, we will show that they exponentiate into a simple Sudakov, similarly to leading power. 

\subsection{General Structure of Factorization and Consistency Relations}\label{sec:consistency_n1}

The complete set of contributions required to perform subleading power resummation involve kinematic corrections, subleading power hard scattering operators, and radiative corrections arising from the subleading Lagrangians in the effective theory. The treatment of these different contributions was worked out in detail in \cite{Moult:2019mog}, and operator bases for both Higgs decays and $e^+e^-$ annihilation were derived in \cite{Moult:2017rpl,Feige:2017zci,Chang:2017atu}. The resummation of the leading logarithms for thrust in pure gluo-dynamics considered in \cite{Moult:2018jjd} used only a subset of these operators. Here we are able to illustrate the complete set of operators required for the resummation of leading logarithms with both quarks and gluons. We believe that describing their structure all together illustrates a rigid and remarkably simple structure imposed by consistency. Furthermore, as has been emphasized above, apart from endpoint contributions, the same set of operators, renormalization and mixings are also sufficient in QCD. To simplify the discussion, we ignore in our presentation contributions arising from kinematic corrections, which were treated in \cite{Moult:2018jjd}, and focus only on those involving non-trivial additional fields (However, contributions from the kinematic corrections are reinstated in the final answer).

Subleading power operators in SCET involve additional quark or gluon fields. Renormalization group consistency implies that for every operator structure with an additional collinear field, there is an analogous operator with an additional soft field, and so terms in the factorization necessarily appear in pairs. This splitting into two contributions arises simply from the fact that in SCET one has factorized into separate fields describing the soft and collinear dynamics, i.e. there are separate soft and collinear quark fields, and separate soft and collinear gluon fields. Indeed, it is apparent from the structure of the matrix elements involved that these two components are describing respectively the soft and collinear limits of the same full theory diagrams. This distinction works somewhat differently at leading power for hard scattering processes, since soft quark and gluon fields do not appear explicitly (soft gluons arise only through Wilson lines from BPS decoupling).

The complete description of all operators requires the introduction of a fair amount of technical machinery and notation.\footnote{The interested reader can find the technical details about SCET hard scattering operators in \cite{Moult:2017rpl,Feige:2017zci,Chang:2017atu}, subleading Lagrangian insertions in \cite{Moult:2019mog} and the structure of subleading power renormalization group evolution in \cite{Moult:2018jjd}.}
Instead of distracting the reader with these details, here we choose to describe the basic structure of the factorization pictorially, since it in fact takes a simple and intuitive structure. For those familiar with SCET, it should be a simple exercise to translate these beautiful pictures into equations. We also provide a more detailed analysis of the ingredients entering the factorization in \app{fact} and we will describe the renormalization of the operators in \Sec{sec:quark_n1}.

In QCD at LL order, renormalization group consistency is separately obeyed for two categories of contributions, that we refer to as quark and gluon corrections. Up to NNLO the utility of separating the calculation into these two categories was first noted in~\cite{Moult:2016fqy}. 
To see why these two categories are distinct for the LL terms to all orders in $\alpha_s$, we first note that at $n$'th loop order they can be computed from contributions with $n-1$ hard loops and $1$ collinear phase space integral, as indicated in \eq{constraints_final}.  In SCET these terms are generated by two distinct categories of operators, either a sub-subleading power operator interfered with a leading power operator for the gluon corrections, $O^{(2)}O^{(0)}$, or  the square of a subleading power operator ${\cal O}^{(1)} {\cal O}^{(1)}$ for the quark corrections. In each case there is a unique operator structure for ${\cal O}^{(1)}$ and ${\cal O}^{(2)}$ that contributes for the LL terms~\cite{Moult:2017rpl,Feige:2017zci}. Since the hard loops correspond to corrections that can be computed at the amplitude level, there is no mixing between the two categories, since the operators appear at different orders in the power counting.\footnote{We also note that the ${\cal O}^{(1)}$ subleading quark operator has a different fermion number in the $n$-collinear sector than the $O^{(0,2)}$ operators contribution to the gluon category, and hence their hard Wilson coefficients are not connected by reparameterization invariance.} 
In addition, for each  category there is also a contribution from subleading power soft interactions, involving a soft gluon for the gluon category, and a soft quark for the quark category. Again at LL order they come from unique operator structures involving two distinct types of Lagrangian insertions for $e^+e^-$ annihilation (wth a quark current)~\cite{Moult:2019mog}, or a Lagrangian insertion and hard scattering operator for Higgs decay (with a gluon current)~\cite{Moult:2017rpl,Moult:2019mog}. 
Below we will separate these two
distinct categories of corrections by square brackets.  
The structure of these pairs is effectively identical for the Higgs decays and $e^+e^-$ annihilation, up to trivial exchanges.
This gives the factorization formula at subleading a fairly transparent structure, consisting of pairs of operators correcting either the gluon or quark emission.

For the case of $e^+e^-$ annihilation, the factorization can be depicted as
\begin{align}\label{eq:factorization_ee}
&\frac{1}{\sigma_0}\frac{d\sigma^{(2),e^+e^-}}{d\tau}=  \\
&\underbrace{\left [ \left| \fd{1.5cm}{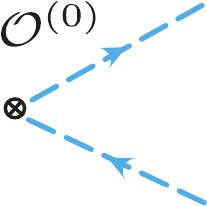} ~ \right|^2\cdot\int d r_2^+ \fd{3cm}{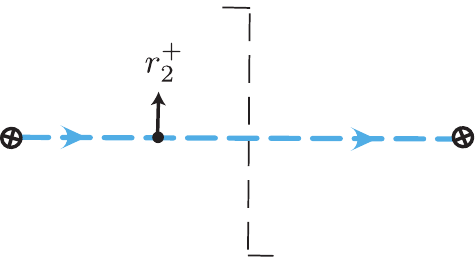} \otimes \fd{3cm}{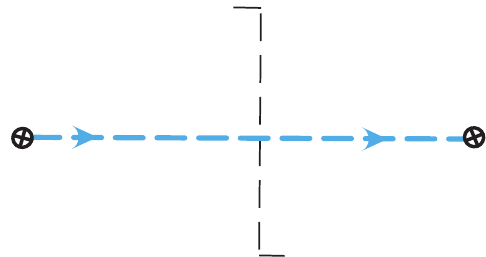}\otimes  \fd{3cm}{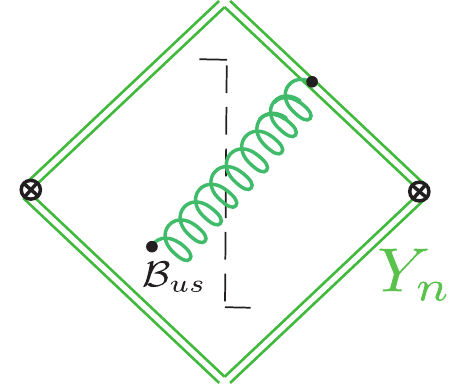} \right. }_{\text{Soft Gluon Correction}}\nn \\
+&\underbrace{ \left. \int d\omega_1 \Re\left ( \fd{1.5cm}{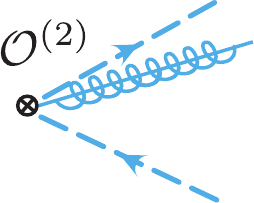}\cdot \fd{1.5cm}{figures_b/2quark_hardfunc_low.pdf}^\dagger \right )  \otimes\fd{3cm}{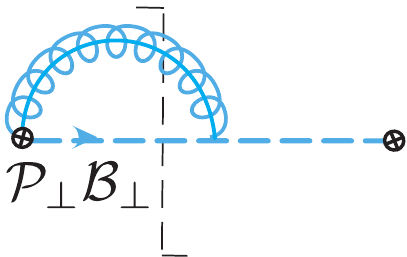}\otimes \fd{3cm}{figures_b/1quark_jetfunction_low} \otimes \fd{3cm}{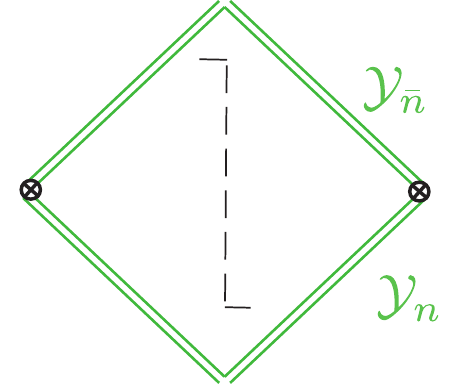} \right ]}_{\text{Collinear Gluon Correction}}\nn \\
+&\underbrace{ \left [  \left| \fd{1.5cm}{figures_b/2quark_hardfunc_low.pdf}~ \right|^2  \cdot   \int dr_2^+ dr_3^+ \fd{3cm}{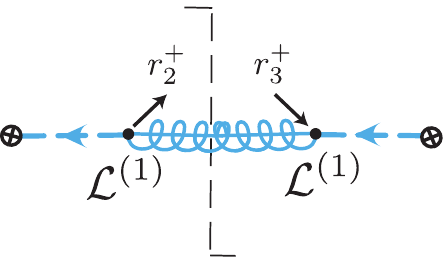}\otimes \fd{3cm}{figures_b/1quark_jetfunction_low} \otimes  \fd{3cm}{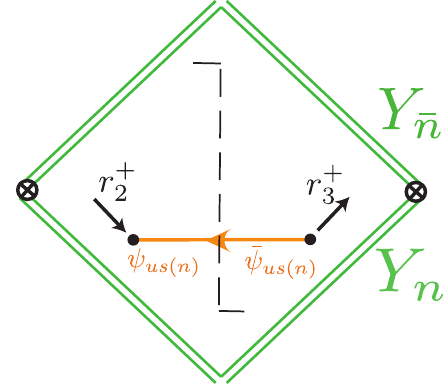}  \right. }_{\text{Soft Quark Correction}}\nn \\
 +&\underbrace{ \left. \int d \omega_1 d \omega_2 \left| \fd{1.5cm}{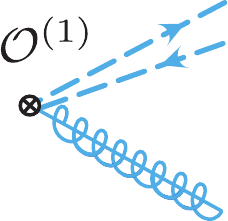}~ \right|^2\otimes\fd{3cm}{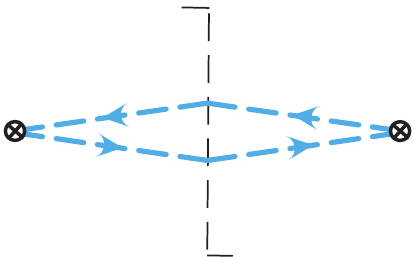}\otimes \fd{3cm}{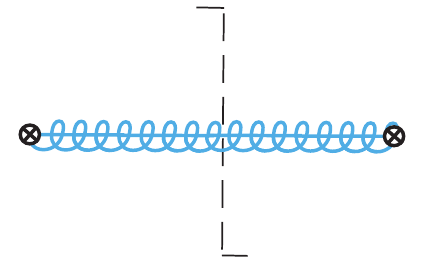}\otimes \fd{3cm}{figures_b/eikonal_factor_gluon_low.pdf} \right]}_{\text{Collinear Quark Correction}}\,. \nn
\end{align}
In each line, we show from left to right the hard function, then the jet functions, and finally the soft function. Convolutions in the observable, $\tau$ are suppressed, and are incorporated in the $\otimes$ symbols. Convolution variables in the field momenta, which appear in subleading power factorization are made explicit.
This factorization involves two separate RG invariant categories, which are indicated by the square bracket. 
The first category describes the corrections to the emission of a soft gluon, which are incorporated through the addition of a gauge invariant soft gluon field, or a gauge invariant collinear field. 
In either case, there is a single variable convolution appearing in the factorization, which describes a light cone momentum of the given field. 
These appear as the interference of an $\cO(\lambda^2)$ operator or Lagrangian insertion with a $\cO(\lambda^0)$ operator. 
This is forced by the LBK theorem \cite{Low:1958sn,Burnett:1967km}, which guarantees that there is no $\cO(\lambda)$ correction for soft gluon emission at amplitude level. 
The second category describes the corrections to quark emission, which are incorporated through the addition of a gauge invariant soft quark field, or a gauge invariant collinear quark field. Due to fermion number conservation, these fields must be inserted on either side of the cut. 
From a technical perspective, they therefore appear as the interference of two $\cO(\lambda)$ operators, or two $\cL^{(1)}$ Lagrangian insertions in the effective theory. This leads to a factorization structure with two integration variables. A more detailed analysis of the ingredients entering in \eq{factorization_ee} can be found in \app{app_factorization_ee}.

For the case of Higgs decay, the factorization can be depicted as
\begin{align}\label{eq:factorization_Higgs}
&\frac{1}{\sigma_0}\frac{d\sigma^{(2),\text{Higgs}}}{d\tau}= \\
&\underbrace{\left[~ \left|\fd{1.8cm}{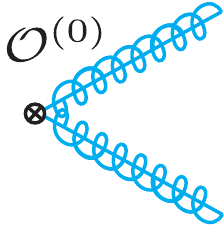}~ \right|^2\,\cdot \, \fd{3.3cm}{figures_b/1gluon_jetfunc_low}\,\otimes \,\fd{3.3cm}{figures_b/1gluon_jetfunc_low}\, \otimes \, \fd{3cm}{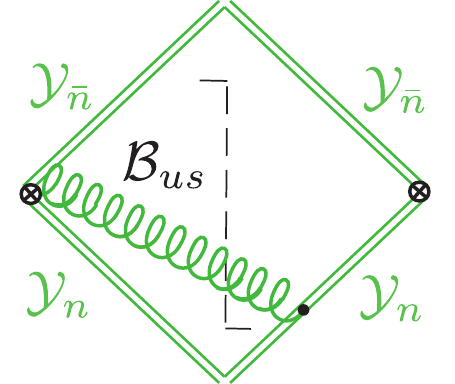}  \right.}_{\text{Soft Gluon Correction}} \nn \\
&+ \underbrace{ \left. \int d\omega_1 \Re \left (\fd{1.5cm}{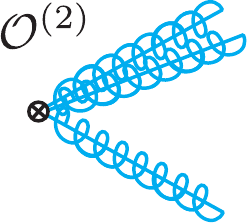}  \cdot \fd{1.5cm}{figures_new/2gluon_hard_low.pdf}^\dagger  \right) \otimes\fd{3cm}{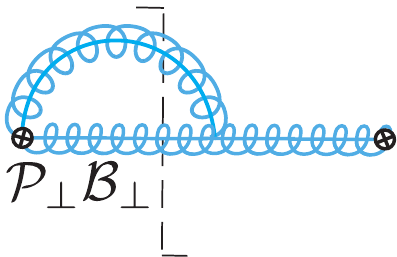}\otimes \fd{3cm}{figures_b/1gluon_jetfunc_low} \otimes \fd{2.7cm}{figures_b/eikonal_factor_gluon_low.pdf} \right ] }_{\text{Collinear Gluon Correction}}   \nn \\
&+ \underbrace{ \left[   \left| \fd{1.5cm}{figures_new/2gluon_hard_low.pdf}  ~\right|^2 \cdot   \int dr_2^+ dr_3^+ \fd{3cm}{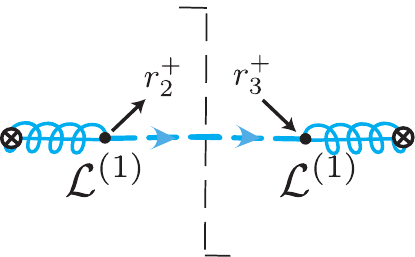}\otimes \fd{3cm}{figures_b/1gluon_jetfunc_low} \otimes  \fd{3cm}{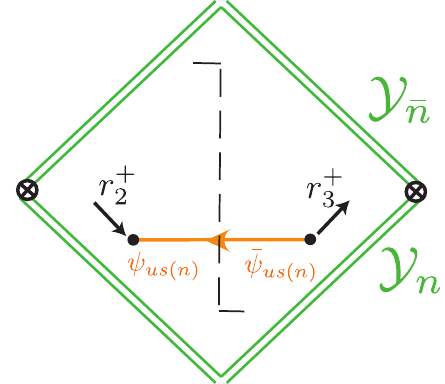} \right . }_{\text{Soft Quark Correction}}\nn \\
&+ \underbrace{ \left. \int d \omega_1 d \omega_2  \left| \fd{1.5cm}{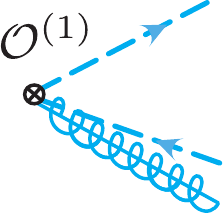}~ \right|^2  \otimes\fd{3cm}{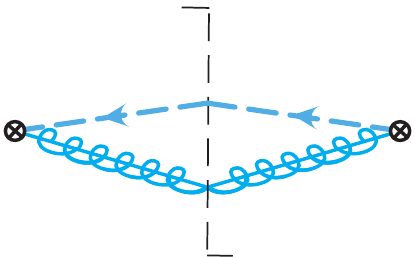}\otimes \fd{3cm}{figures_b/1quark_jetfunction_low}\otimes \fd{3cm}{figures_b/eikonal_factor_gluon_low.pdf} \right ]}_{\text{Collinear Quark Correction}}\,.\nn
\end{align}
This takes an essentially identical structure to the case of $e^+e^-$ annihilation, up to the exchange of some quark and gluon fields. Indeed, in $\cN=1$ one finds an identical result in the two cases. A more detailed analysis of the ingredients entering in \eq{factorization_Higgs} can be found in \app{app_factorization_Higgs}.

While it may naively seem like there are a large number of operators involved in this factorization, we see that it ultimately has a simple structure: the four different contributions are understood as corrections to quarks and gluons that are then split between soft and collinear fields. With this structure understood, and using the formalism for the renormalization of subleading power operators that was introduced in \cite{Moult:2018jjd}, and will be reviewed in \Sec{sec:quark_n1}, LL resummation at subleading power in the $\cN=1$ theory can be performed almost as easily as at leading power.

\subsection{Renormalization and Leading Logarithmic Resummation}\label{sec:quark_n1}

In \cite{Moult:2018jjd}, an understanding of the renormalization of subleading power operators was developed (working in a case where no endpoint divergences are present, as is the case at LL in $\cN=1$ SUSY). Renormalization is conceptually non-trivial for subleading power soft and jet functions that involve additional fields (for example the soft functions involving an additional soft quark or gluon field, or the jet functions involving an additional collinear quark or gluon field that were shown in \Eqs{eq:factorization_ee}{eq:factorization_Higgs}). The renormalization of these operators is described by mixing into a new set of subleading power operators, which were termed $\theta$-jet and $\theta$-soft functions, that are not seen in the amplitude level matching. In \cite{Moult:2018jjd} it was shown using renormalization group consistency that under general assumptions related to the convergence of convolution integrals, the matrix elements of these new subleading power operators are given by moments of the leading power functions. 

The $\theta$-soft function is defined as
\begin{align}\label{eq:theta_soft_first}
S^{(2)}_{g,\theta}(k,\mu)&= \frac{1}{(N_c^2-1)} \tr \langle 0 | \cY^T_{\bar n} (0)\cY_n(0) \theta(k-\hat \Tau) \cY_n^T(0) \cY_{\bar n}(0) |0\rangle\,,
\end{align}
while the $\theta$-jet functions for quarks and gluons are defined as
\begin{align}\label{eq:theta_op}
J^{(2)}_{g,\theta}(s,\mu)&=\frac{(2\pi)^3}{(N_c^2-1)} \Big\langle 0 \Big|\, \cB^{\mu a}_{n\perp} (0)\,\delta(Q+\bar \cP) \delta^2(\cP_\perp)\, \theta\left(\frac{s}{Q}-\hat \Tau\right)\, \cB^{\mu a}_{n\perp,\omega}(0) \,\Big|0\Big\rangle\,,  \\
J^{(2)}_{q,\theta}(s,\mu)&=\frac{(2\pi)^3}{N_c} \Big\langle 0 \Big|\, \bar \chi_{n} (0)\,\delta(Q+\bar \cP) \delta^2(\cP_\perp)\, \theta\left(\frac{s}{Q}-\hat \Tau\right)\, \chi_{n,\omega}(0) \,\Big|0\Big\rangle\,.\nn
\end{align}
These functions obtain their power suppression from the $\theta$ function, which is power suppressed relative to the standard $\delta$ function that appears in jet and soft functions.
They obey the RG equations
\begin{align}\label{eq:theta_RG}
\mu \frac{d}{d\mu}  S^{(2)}_{g,\theta}(k,\mu)& = \int dk' \gamma^S_g(k-k')S^{(2)}_{g,\theta}(k',\mu)\,, \\
\mu \frac{d}{d\mu}  J^{(2)}_{g,\theta}(s,\mu) &= \int ds' \gamma^J_{g}(s- s')J^{(2)}_{g,\theta}(s',\mu)\,, \\
\mu \frac{d}{d\mu}  J^{(2)}_{q,\theta}(s,\mu) &= \int ds' \gamma^J_{q}(s- s')J^{(2)}_{q,\theta}(s',\mu)\,.
\end{align}
Here $\gamma^S_g$, $\gamma^J_{g}$ and $\gamma^J_{q}$ are the anomalous dimensions of the leading power soft and jet functions, which can be found in \cite{Moult:2018jjd} .

With this understanding of the structure of the renormalization and factorization at subleading power, it is now a technically straightforward exercise to derive the LL resummed result at subleading power. Using renormalization group consistency, we can choose to run all operators to the soft scale. This eliminates the need to consider the soft functions involving radiative quarks or gluons.  We therefore must simply renormalize the hard functions involving two collinear quark fields, or a quark and gluon collinear field, as well as the subleading power jet functions. 

Each of the jet functions involving multiple fields (which we denote generically as $J^{(2)}_{j}$) obeys a two-by-two mixing equation, in which they mix with an adjoint $\theta$-jet function (either quark or gluon). This mixing takes the form
\begin{align}\label{eq:2by2mix}
\mu \frac{d}{d\mu} \Biggl(\begin{array}{c} J^{(2)}_{j}(s,\mu) \\[5pt]
  J^{(2)}_{g/q,\theta}(s,\mu) \end{array} \Biggr) &= \int ds' 
  \Biggl( \begin{array}{cc} 
\gamma^J_{jj}(s-s') \qquad &  \gamma^J_{j\theta}\, \delta(s-s') \\[5pt]
  0 &  \gamma^J_{g/q, \theta \theta}(s-s')    
\end{array} \Biggr) \,,
\end{align}
at leading logarithm. Note that we have supressed the dependence on the momentum labels of the fields, since at leading log the renormalization is diagonal in these momentum labels. 
This equation was solved in \cite{Moult:2018jjd} with and without a running coupling. Without running coupling, it leads to a standard Sudakov exponentiation.  Since the running coupling provides no additional insight to the structure at subleading power, we will not bother to consider it here.

From explicit calculation, at LL each of the different hard functions in N=1 SUSY QCD satisfy a multiplicative RG equation determined by the standard cusp anomalous dimensions
\begin{align}
\mu \frac{d}{d\mu}H_{j} (\omega_1, \omega_2, Q^2, \mu^2)&= \Gamma_\cusp[\alpha_s(\mu)] \log \left( \frac{-Q^2}{\mu^2}  \right)  H_{j} +\cdots\,,
\end{align}
where the ellipses denote non-cusp terms that do not contribute to the leading logarithmic result. Here the subscript $j$ is supposed to denote the different field configurations appearing in the hard functions in \Eqs{eq:factorization_ee}{eq:factorization_Higgs}.  Note that since we are in $\cN=1$, the different cusp anomalous dimensions are all the same, $ \Gamma^g_\cusp= \Gamma^q_\cusp$, which enables us to treat all the anomalous dimensions on the same footing.

Once it is shown that the renormalization of the operators leads to a standard Sudakov exponentiation, it is then a simple combinatorial exercise to add up the contributions from different operators. We spare the reader these details, and just give the final result. We find that the NLP results for both Higgs decay and $e^+e^-$ give an identical (as expected in $\cN=1$) and simple result
\begin{align}\label{eq:N1_resummed}
\frac{1}{\sigma_0} \frac{d\sigma_{\text{LL}}^{(2),\text{Higgs}}}{d\tau}&=\frac{1}{\sigma_0} \frac{d\sigma_{\text{LL}}^{(2),e^+e^-}}{d\tau}=\left(  \frac{\alpha_s}{4\pi} \right) 4C_A \log(\tau) e^{-4C_A\left(  \frac{\alpha_s}{4\pi} \right)  \log^2( \tau)}\,.
\end{align}
Note that this agrees with the fixed order results given in \cite{Moult:2016fqy,Moult:2017jsg} upon taking $C_F\to C_A$ and $n_f\to C_A$ to obtain $\cN=1$ from QCD.

We believe that this result is interesting in its own right, since it shows that resummation of subleading power corrections at leading logarithm is understood in $\cN=1$ (as well as for theories with more supersymmetry), and indeed has a simple form that, once understood, is technically no more difficult than LL resummation at leading power. In particular, we have  illustrated that the mixing with $\theta$ function operators first introduced in \cite{Moult:2018jjd} applies much more generally.  We hope that the fact that factorization works in this manner in $\cN=1$ may provide clues for how it can be extended to QCD. In particular, it seems to indicate that the only issue in QCD arises from the $z\to 0$ or $z\to 1$ limits of the convolution. 

\section{A Conjecture for the Soft Quark Sudakov in QCD}\label{sec:soft_quark_sudakov}

Although we are not yet in a position to derive the all loop result for soft quark emission in QCD from an operator level factorization, our explicit calculation to $\cO(\alpha_s^3)$, combined with insight from the exponentiation in $\cN=1$ QCD allows us to make a conjecture for the result. It will be interesting to derive this result (or prove that it is incorrect) from an operator level renormalization and factorization analysis in the future.

We arrive at our conjecture for the LL Sudakov in QCD by assuming that the radiative corrections to the squared amplitude exponentiate (at least for the highest order poles in $\epsilon$). Namely, that
\begin{align}\label{eq:Pijexponen}
	P_{ij}^{(2,n)}(z,\epsilon)=P_{ij}^{(2,0)}(z) \frac{K(z,\epsilon)^n}{n!} + \cO(1/\epsilon^{2n - 1})\,,
\end{align}
where $K(z,\epsilon)$ describes the leading infrared poles in $\epsilon$ 
\begin{align}\label{eq:Kdef}
	K(z,\epsilon) =&  \frac{\alpha_s}{4\pi} \mu^{2\epsilon} \Big[ 4 {\bf{T}}_1 \cdot {\bf{T}}_2    \frac{ [(1-z)Q^2]^{-\epsilon}}{\epsilon^2} +4 {\bf{T}}_1 \cdot {\bf{T}}_3 \frac{s^{-\epsilon}}{\epsilon^2}  +4 {\bf{T}}_2 \cdot {\bf{T}}_3 \frac{[zQ^2]^{-\epsilon}}{\epsilon^2}  \\
&  +4 {\bf{T}}_1 \cdot {\bf{T}}_3 \left(  \frac{[Q^2]^{-\epsilon}}{\epsilon^2}-  \frac{[z(1-z)Q^2]^{-\epsilon}}{\epsilon^2}  \right) -4 {\bf{T}}_1 \cdot {\bf{T}}_3  \left(  \frac{s^{-\epsilon}}{\epsilon^2}-  \frac{[z(1-z)s]^{-\epsilon}}{\epsilon^2}  \right) \Big] \nn\,.
\end{align}
This result was illustrated in \Sec{sec:nlp_collinear} for different partonic channels. Note that we have explicitly checked this for $n=1,2$ by direct calculation, and we know from our factorization and renormalization analysis that it holds for the cusp contributions, but we do not strictly speaking know that it holds for the endpoint contributions.
This exponentiated  conjecture for the splitting function can then be integrated against the measurement function, as was done in \cite{Moult:2018jjd} to derive the result for a physical observable using the consistency equations in \Sec{sec:consistency}.

One of the reasons we make a conjecture for the result, is that under this assumption for the exponentiation of the splitting function, we find after integration, as we show in \app{PSintegral}, that the color structure at any order is given by
\begin{align}\label{eq:sym_color}
C_A^n + C_A^{n-1}C_F +\cdots C_A C_F^{n-1}+ C_F^n\,.
\end{align}
This is not at all apparent before integration. Intriguingly, identical color structures have appeared in the $x \to1$ resummation of the off diagonal splitting functions \cite{Vogt:2010cv,Almasy:2010wn,Presti:2014lqa,Almasy:2015dyv}. We believe that this is not a coincidence, and we believe that it provides some non-trivial evidence for our conjecture.  Alternatively, if one can argue for the color structure \Eq{eq:sym_color}, as was done in \cite{Vogt:2010cv,Almasy:2010wn,Presti:2014lqa,Almasy:2015dyv} for the off-diagonal splitting functions, then one can extend the result in $\cN=1$, where it can be proven to the case of QCD by adding terms proportional to $(C_A-C_F)^n$.

Using either line of reasoning, both of which are conjectural, we can make an ansatz for the full result for the subleading power leading logarithmic result for thrust in $e^+e^-$ in full QCD
\begin{align}\label{eq:conjecture_ee}
\frac{1}{\sigma_0} \frac{d\sigma_{\text{LL}}^{(2),e^+e^-}}{d\tau}&=\left(  \frac{\alpha_s}{4\pi} \right) 8C_F \log(\tau) e^{-4C_F \left(  \frac{\alpha_s}{4\pi} \right)  \log^2( \tau)} \nn \\
&+\underbrace{\frac{C_F}{(C_F-C_A)\log(\tau)}  \left(   e^{-4C_F \left(  \frac{\alpha_s}{4\pi} \right)  \log^2( \tau)}   - e^{-4C_A \left(  \frac{\alpha_s}{4\pi} \right)  \log^2( \tau)}  \right)}_{\text{Soft Quark Sudakov}}\,.
\end{align}
Here the first term arises from the subleading power corrections to soft gluon emission, which take the form of a simple Sudakov, as shown in \cite{Moult:2018jjd}, and the second term gives the conjectured ``soft quark Sudakov", which takes a more non-trivial form. Despite this more unusual structure, the result still seems to be controlled by a difference of cusp anomalous dimensions, as expected from the physical picture developed in \Sec{sec:factorization}. This suggests that understanding the endpoint divergences is a conceptual rather than technical difficulty.

We can write this soft quark Sudakov in a more transparent form, as a modification of the standard Sudakov by $C_A-C_F$ terms
\begin{align}
\text{Soft Quark Sudakov}=-4 C_F  \left(  \frac{\alpha_s}{4\pi} \right) \log(\tau) e^{-4C_F \left(  \frac{\alpha_s}{4\pi} \right)  \log^2( \tau)}    \left[   \frac{  \left( 1-  e^{-4(C_A-C_F) \left(  \frac{\alpha_s}{4\pi} \right)  \log^2( \tau)}  \right)}{4(C_A-C_F) \left(  \frac{\alpha_s}{4\pi} \right)\log^2(\tau)}  \right]\,.
\end{align}
Here the term in square brackets becomes 1 in the $C_F\to C_A$ limit. It can be viewed as dressing the standard Sudakov that appears in the $\cN=1$ theory. It represents a new all orders structure first appearing at subleading power. Functions of $(C_A-C_F)$ have also been observed in the high-energy limit for mass suppressed amplitudes in \cite{Liu:2017vkm,Liu:2018czl}. It would be interesting to understand the connections between these results in more detail.

For the case of thrust in Higgs decays, one can conjecture a similar result
\begin{align}\label{eq:conjecture_H}
\frac{1}{\sigma_0} \frac{d\sigma_{\text{LL}}^{(2),\text{Higgs}}}{d\tau}&=\left(  \frac{\alpha_s}{4\pi} \right) 8C_A \log(\tau) e^{-4C_A \left(  \frac{\alpha_s}{4\pi} \right)  \log^2( \tau)} \nn \\
&+\frac{n_f}{(C_F-C_A)\log(\tau)}  \left(   e^{-4C_F \left(  \frac{\alpha_s}{4\pi} \right)  \log^2( \tau)}   - e^{-4C_A \left(  \frac{\alpha_s}{4\pi} \right)  \log^2( \tau)}  \right)\,.
\end{align}
Note that setting $C_F,~n_f \to C_A$ in either \Eq{eq:conjecture_ee} or \Eq{eq:conjecture_H} reproduces the $\cN=1$ formula of \Eq{eq:N1_resummed}.

It is worth emphasizing that since this result follows from consistency relations, combined with the conjectural structure of the amplitudes, it will have considerable universality. In particular, it should hold for any \SCETi type dijet observable (recoil free observable) up to trivial color factor changes. For this reason we refer to it as ``the" soft quark Sudakov. 

One numerically interesting feature of this result is that for a process that has a Sudakov in a given representation at leading power, at subleading power, both representations appear. Since the Sudakov with a $C_A$ dies off more rapidly, in $e^+e^-$, these corrections have a small effect. On the other hand, the effect for Higgs decays is much larger, since there is now a Sudakov with a $C_F$, which dies off more slowly.

It is also interesting that while at leading power and leading logarithm, the results for $\cN=1$ QCD and QCD are identical up to a color factor, at subleading power they have a significantly different form. Interesting simplifications in the infrared limits of $\cN=1$ were also observed in collinear limits in \cite{Dixon:2019uzg} and at two loop level for the $2\to 2$ amplitude in $\cN=2$ in \cite{Duhr:2019ywc}. This suggests that further understanding at subleading powers may lead to a better understanding of the differences in their infrared and transcendental structure.

\section{Conclusions}\label{sec:conc}

In this paper we have initiated a study of the all order loop level structure of soft fermion emission, which generically contributes leading infrared logarithms at subleading power. This supplements studies of subleading power corrections to soft gluon emission which have 
been resummed using the formalism introduced in \cite{Moult:2018jjd}.

We computed the subleading power collinear limits of $e^+e^-\to 3$ parton and $H\to 3$ parton amplitudes at two loops, focusing in particular on limits that do not have a leading power analogue. We found that in the case of QCD, Sudakov exponentiation is violated by endpoint divergences. These endpoint contributions were shown to be proportional to $(C_A-C_F)^n$, and hence vanish in $\cN\geq 1$ QCD. We also gave a physical picture for  these endpoint contributions, which we believe are related to a non-trivial zero bin when $C_A\neq C_F$. We leave finding a operator renormalization based understanding of the structure of these terms to future work. 

We performed an operator level factorization and resummation of the infrared logarithms associated with soft quark emission in $\cN=1$ QCD. This extends our analysis in \cite{Moult:2018jjd} to include new subleading power jet and soft functions involving quarks. The renormalization of these operators involves mixing into $\theta$ function operators as was found for the pure gluon operators considered in \cite{Moult:2018jjd}. Using this analysis, we showed that radiative contributions associated with soft quark emissions exponentiate into a Sudakov exponential. This also shows that leading logarithmic resummation at subleading powers is now understood in $\cN\geq 1$ QCD.

Using the results from our fixed order analysis, as well as the structure of the exponentiation in $\cN=1$ QCD, we made a conjecture for the all loop structure of the soft quark Sudakov in QCD. We believe that this structure will govern all \SCETi dijet event shapes, and we have provided some non-trivial evidence of this universality by considering the event shape thrust for both Higgs decay and $e^+e^-$ annihilation.  It would be extremely interesting to prove (or disprove) this formula by better understanding the endpoint contributions, as well as to understand its relation to the off-diagonal splitting functions. 

An interesting direction that deserves further study is to extend this work to the case of perpendicular momentum sensitive observables. These observables are affected by rapidity divergences and obey a set of additional rapidity renormalization group equations that are well understood at leading power \cite{Chiu:2012ir}. The study of rapidity divergences at subleading power has been initiated at fixed order in \cite{Ebert:2018gsn}. While we believe that most of the structures we found in this current work will be present also for observables with rapidity divergences, the structure of the rapidity renormalization group at subleading power remains to be understood.

We believe it is of utmost interest to understand the structure of endpoint divergences in subleading power factorization, as this is now the primary obstacle to a complete understanding. The fact that the only difference in the LL subleading power resummation between $\cN=1$ SUSY and QCD resides in endpoint contributions indicates that they are not simply a technical detail, but describe the interesting physical effect of having particles in different color representations. We find it interesting that the subleading power limit is so sensitive to the particular representations and matter content of the theory, and we therefore hope that better understanding subleading power limits can lead to an improved understanding of the infrared behavior of amplitudes and cross sections in gauge theories.

\begin{acknowledgments}
We thank Alexander Penin, Bernhard Mistlberger, and Duff Neill for interesting conversations. We thank Martin Beneke, Robert Szafron and Sebastian Jaskiewicz for many interesting discussions related to subleading power factorization and resummation. We thank the participants of the 2018 Nikhef ``Next to leading power corrections in particle
physics'' workshop for stimulating discussions. We thank DESY and the MIT Center for Theoretical Physics for hospitality while portions of this work were performed. 
This work was supported in part by the Office of Nuclear Physics of the U.S.
Department of Energy under Contract No. DE-SC0011090, by the Office of High Energy Physics of the U.S. Department of Energy under Contract No. DE-AC02-76SF00515, and by the Simons Foundation Investigator Grant No. 327942.
\end{acknowledgments}

\appendix
\section{From subleading splitting functions to thrust power corrections}\label{app:PSintegral}
In this appendix we want to show how power corrections for a recoil free observable, like thrust, arise from the collinear expansion of amplitudes. For concreteness, we will focus on the case of thrust in $e^+e^- \to $ jets and analyze how the subleading power splitting amplitudes $P^{(i,j)}_{q\bar q}$ of \eqs{qg_collinear}{qg_collinear_2} give rise to the second line of \eq{conjecture_ee} at the cross section level, and in particular to the peculiar color structure in \eq{sym_color}.
In doing so we take the occasion to point out some differences between these $P^{(i,j)}_{q\bar q}$ and other expansions of splitting functions.

In \sec{consistency_n1} we have shown that we can extract the leading logarithmic contribution at ${\cal O}(\alpha_s^{n+1})$ from the $n-$loop correction of the single real collinear emission. 
Following \eq{fig_ee}, let us define $P_{q\bar q}(\tau,z,\epsilon)$ as the amplitude squared for $e^+e^- \to 3 $ partons in $d=4-2\epsilon$ dimensions, which is differential in both the momentum fraction $z$ of the quark and in the rescaled invariant mass $\tau=(p_1 + p_2)^2/Q^2$ of the quark-antiquark pair. At fixed $z$, the $\tau\to0$ limit singles out the configurations where the quark anti-quark pair become collinear.
$P_{q\bar q}(\tau,z,\epsilon)$ can be expanded%
\footnote{Note that in general $P_{q\bar q}(\tau,z,\epsilon)$ will have terms proportional with non integer powers of $\tau$, like $\tau^{\epsilon}$, see for example the last term of the second line of \eq{Pqq21} with $s\equiv Q^2\tau$. In the present analysis, where we want to get the leading log behavior via consistency just from the hard loop corrections, we can ignore  these $\epsilon$-onic powers since they are associated with contributions arising from the collinear region of the loop momentum.}
in $\tau$
and $\alpha_s$
\be\label{eq:Pqqexpansion}
	P_{q\bar q}(\tau,z,\epsilon) = \frac{1}{\tau}\sum_{i,j}\tau^i \left(\frac{\alpha_s}{4\pi}\right)^j P_{q\bar q}^{(2i,j)}(z,\eps)\,. 
\ee
Note that the subleading splitting amplitude $P^{(i,j)}_{q\bar q} (z,\epsilon)$ is a non trivial function of $z$. When we say \emph{subleading power} or \emph{next-to-leading power} in the expansion, this refers only to the expansion in $\tau$, while the dependence on $z$ is kept at all orders.
This is in particular clear if we already look at the tree level subleading splitting amplitude $P_{q \bar q}^{(2,0)}(z,\epsilon)$ in \eq{Pqq20}. $P_{q \bar q}^{(2,0)}(z,\epsilon)$ not only contains an all order dependence on $z$, but it also contains singularities both as $z\to0$ and as $z\to1$. Since we are studying the logarithms of $\tau$ after integrating over $z$, this differs from the study in Refs.~\cite{Vogt:2010cv,Soar:2009yh} of the off-diagonal splitting functions themselves which is closely related to the study of threshold logarithms at subleading power from the expansion about the limit $z\to 1$, rather than those for thrust. In general these are simply different observables, and one does not expect a simple relation between their resummed form for the soft quark channel at subleading power. Even at leading power these observables differ when one considers the form of their higher order resummed formula.

Next we show that at the cross section level, the color structure of \eq{sym_color} arises.
For concreteness, we focus on the subleading splitting amplitude at ${\cal O}(\alpha_s^3)$, which corresponds to $P_{q\bar q}^{(2,3)}(z,\eps)$. As explained in \sec{consistency_n1} and in \eq{constraints_final}, at order $\alpha_s^{n}$ the leading logarithmic contribution to the NLP $\tau$-dependent cross section, $\alpha_s^n \log^{2n-1} \tau$, can be obtained by consistency via the coefficients $c_{hc,2n-1}$, which are the prefactors for the most singular $1/\epsilon^k$ divergence at this order in $\tau$. 
In the following we will calculate $c_{hc,5}$, which is obtained by integrating the subleading splitting amplitude at $2$-loops against the leading power single emission collinear phase space
\begin{align}\label{eq:}
	c_{hc,5} = 
	\Bigg[\int_0^1 \underbrace{\frac{\df z}{z^\epsilon(1-z)^\epsilon}}_{\substack{\text{Leading power} \\\text{two-particle}\\ \text{collinear PS}}} \underbrace{P^{(2,2)}_{q\bar q} (z,\epsilon)}_{\substack{\text{Subleading collinear}\\ \text{amplitude}}}\Bigg]_{\cO\left(\frac{1}{\epsilon^{5}}\right)}\,.
\end{align}
Since the LP splitting amplitude for this quark emission channel vanishes, this is the only contribution and we do not have to consider the LP power splitting amplitude integrated against the NLP phase space.

Because of charge conjugation, we have 
\be 
	P^{(2,2)}_{q\bar q} (z,\epsilon)=P^{(2,2)}_{q\bar q} (1-z,\epsilon)\,.
\ee
Therefore,
\begin{align}\label{eq:chc5}
	\frac{c_{hc,5}}{\epsilon^5} &= 
	2\int_0^{1/2} \frac{\df z}{z^\epsilon}P^{(2,2)}_{q\bar q} (z,\epsilon) =2\int_0^{1/2} \frac{\df z}{z^\epsilon}P^{(2,0)}_{q\bar q} (z,\epsilon)\frac{K^2(z,\epsilon)}{2!}\,.
\end{align}
Plugging in the expression for $P^{(2,0)}_{q\bar q}$ from \eq{Pqq20}, and $K(z,\epsilon)$ from \eq{Kdef}, translating the color generators in terms of Casimirs from \eq{Casimirsee}, and keeping only the terms relevant for the required pole, we have
\begin{align}\label{eq:chc5}
	\frac{c_{hc,5}}{\epsilon^5}  &= 
\int_0^{1/2} \frac{\df z}{z^\epsilon}\frac{4C_F}{z}\,\frac{1}{2!}\left[\frac{4C_F +4(C_A-C_F)z^{-\epsilon}}{\epsilon^2}\right]^2 
 \\&
 =\frac{32C_F}{\epsilon^4}\int_0^{1/2} \frac{\df z}{z^{1+\epsilon}}\,\left[C_F^2 
 +(C_A-C_F)^2z^{-2\epsilon}+ 2C_F(C_A-C_F)z^{-\epsilon}\right]
 \nn\\&
 =\frac{32C_F}{\epsilon^4}\int_0^{1/2} \df z\left[\frac{C_F^2}{z^{1+\epsilon}}+\frac{2C_F C_A-2C_F^2}{z^{1+2\epsilon}}+\frac{C_A^2-2C_FC_A+C_F^2}{z^{1+3\epsilon}}\right]\,.
  \nn\\&
  =\frac{32 C_F}{\epsilon^4} \int_0^{1/2} \df z\left[
  C_F^2 \left(\frac{1}{z^{1+\epsilon}}-\frac{2}{z^{1+2\epsilon}}+\frac{1}{z^{1+3\epsilon}}\right)
  +2C_F C_A\left(\frac{1}{z^{1+2\epsilon}}-\frac{1}{z^{1+3\epsilon}}\right)
  +C_A^2 \frac{1}{z^{1+3\epsilon}} \right] . \nn
 \end{align}
Now the integration is $z$ is trivial, and after then expanding in $\epsilon$ and keeping the leading pole we obtain
\be
	c_{hc,5}=-\frac{32}{3} C_F \left[
  C_F^2
  +C_F C_A
  +C_A^2\right]\,, \nn
\ee
which matches the form shown in \eq{sym_color} for $n=2$. It is then an easy exercise to show that using the ansatz that $K^n$ exponentiates, one gets \be
	P_{q\bar q}^{(2,LL)}= P_{q\bar q}^{(2,0)} e^{K(z,\epsilon)}\,.
\ee
Using this result, the integration against the leading power two particle collinear phase space then yields the second line of \eq{conjecture_ee}.

\section{Leading Logarithmic Factorization Structure for Thrust at NLP in $\cN=1$}\label{app:fact}

In this appendix we will explain in more detail the ingredients entering the factorization for thrust in $e^+e^-$ annihilation, depicted in \eq{factorization_ee}, and in Higgs decay, depicted in \eq{factorization_Higgs}. Since in the following we are always considering the NLP contribution to the thrust observable, we will use the shorthand notation 
\be
	J \otimes J \otimes S = Q^5 \int\!\! d\tau_n d\tau_{\bar n} d\tau_{us}\: \delta(\tau-\tau_n -\tau_{\bar n}-\tau_{us}) J(Q^2\tau_n)\, J(Q^2\tau_\bn)\, S(Q\tau_{us}) \,.
\ee
At subleading power the soft and jet functions can have additional functional arguments, that are not involved in integrals associated to the observable $\tau$,  and which we will include when they are relevant.
 
\subsection{Higgs decay}\label{app:app_factorization_Higgs}
Let's start with the case of Higgs decay.  In \Tab{tab:higgsops} we list the hard scattering operators contributing at leading logarithm for this process.
{
\renewcommand{\arraystretch}{1.4}
\begin{table}[h]
\begin{center}
\scalebox{0.842}{
\begin{tabular}{| l | c | c |c |c|c| r| }
  \hline                       
   Operator & Tree Level Matching Coefficient \\
  \hline
    $\cO_\cB^{(0)}=C^{(0)} \delta^{ab} \cB_{\perp \bar n, \omega_2}^a \cdot \cB_{\perp \bar n, \omega_1}^b H$& $C^{(0)}=-2\omega_1 \omega_2$ \\
  \hline
  $\cO^{(1)}_{\cB_n}= C^{(1)}_{\cB_n} \left( { Y}_{\bar n}^\dagger {Y}_{n} T^a \right)_{\alpha \bar\beta}  \left (\bar\chi^{ \alpha}_{n, \omega_1} \Sl{\cB}^a_{\bn\perp, \omega_3} \chi^{\bar\beta}_{\bn,- \omega_2}  \right)$ & $C^{(1)}_{\cB_n}=-2g\frac{\omega_3}{\omega_2}$ \\
  \hline
   $\cO^{(2)}_{\cP \cB1}=C^{(2)}_{\cP \cB1}i f^{abc} \cB^a_{n\perp,\omega_1}\cdot \left[  \cP_\perp \cB^b_{\bar n \perp,\omega_2}\cdot  \right] \cB_{\bar n \perp,\omega_3}^c    H$ &  $C^{(2)}_{\cP \cB1}=-\left( \frac{1}{2}\right)4g \left(  2+\frac{\omega_3}{\omega_2}+ \frac{\omega_2}{\omega_3}  \right)$ \\
   \hline
   $\cO^{(2)}_{\cP \cB2}=C^{(2)}_{\cP \cB2} if^{abc}\left[ \cP_\perp \cdot \cB_{\bar n \perp,\omega_3}^a \right] \cB^b_{n\perp,\omega_1} \cdot \cB_{\perp \bar n, \omega_2}^c    H$ & $C^{(2)}_{\cP \cB2}=4g\left( 2+\frac{\omega_3}{\omega_2}  + \frac{\omega_2}{\omega_3}\right)$\\
  \hline  
  $\cO^{(2)}_{\cB(us(n))}=C^{(2)}_{\cB \bn(us)} \left(i  f^{abd}\, \big({\cal Y}_n^T {\cal Y}_{\bar n}\big)^{dc}\right)  \left (  \cB^a_{n\perp, \omega_1} \cdot \cB^b_{\bar n \perp, \omega_2} \bar n \cdot g\cB^c_{us(n)} \right)$ & $C^{(2)}_{\cB \bn(us)}=-2 \omega_2$ \\
  \hline
  $\cO^{(2)}_{\cB(us(\bar n))}= C^{(2)}_{\cB n(us)} \left(i  f^{abd}\, \big({\cal Y}_{\bar n}^T {\cal Y}_{n}\big)^{dc}\right)  \left ( \cB^a_{n\perp, \omega_1} \cdot \cB^b_{\bar n \perp, \omega_2} n\cdot g\cB^c_{us(\bar n)} \right)$ & $C^{(2)}_{\cB n(us)}=-2\omega_1$ \\
  \hline
\end{tabular}}
\end{center}
\caption{
Hard scattering operators that contribute to the LL cross section to $\cO(\lambda^2)$ in gluon induced Higgs decay, along with their tree level matching coefficients. These operators and matching coefficients were derived in \cite{Moult:2017rpl}.
}
\label{tab:higgsops}
\end{table}
}

{
\renewcommand{\arraystretch}{1.4}
\begin{table}[h]
\scalebox{0.842}{
\hspace{0.2cm}\begin{tabular}{| l | c | c |c |c|c| r| }
  \hline                       
  $T$-Product & Example Diagram & Soft Function& Jet Function \\
  \hline
   $\cL^{(1)}\cdot \cL^{(1)}$ & $\fd{3cm}{figures_b/soft_quark_collinear_piece_gluoncase_purjet_low}$ & {\begin{small}$ \langle0|\cY_n  \cY_{\bar n} \bar \psi_{us}(r_2^+) \psi_{us}(r_3^+)\cY_{\bar n} \cY_n |0\rangle$\end{small}} &  {\begin{small}$ \langle0| \cB_{n\perp} \bar \chi_n (r_2^+) \cB_{n\perp}  \chi_n(r_3^+) \cB_{n\perp}  \cB_{n\perp} |0\rangle$  \end{small}} \\
  \hline  
\end{tabular}}
\caption{
Leading order contributions to the $\cO(\lambda^2)$ thrust cross section in gluon induced Higgs decay from radiative functions. The soft quark contribution arises from two $\cL^{(1)}$ insertions.
}
\label{tab:radHiggs}
\end{table}
}

In category 1, i.e. for the gluon corrections, we have two hard scattering contributions:
\begin{align}
&   \overbrace{ \left|\fd{1.5cm}{figures_new/2gluon_hard_low.pdf}~ \right|^2\cdot  \fd{3cm}{figures_b/1gluon_jetfunc_low}\otimes \fd{3cm}{figures_b/1gluon_jetfunc_low} \otimes  \fd{2.5cm}{figures/Bus_softfunction_2_low.pdf}  }^{\text{Soft Gluon Correction}}\nn \\
&= \big|C^{(0)}\big|^2 \cdot J^{(0)}_{g,n} \otimes J_{g,\bn}^{(0)} \otimes S_{\bn B_{us}}^{(2)} + n \leftrightarrow \bn\,,
\end{align}
and 
\begin{align}
& \overbrace{ \int d\omega_1 \Re \left (\fd{1.5cm}{figures_new/collinear_category1_hard_low}  \cdot \fd{1.5cm}{figures_new/2gluon_hard_low.pdf}^\dagger  \right) \otimes\fd{3cm}{figures_b/gg_jetfunction_low}\otimes \fd{3cm}{figures_b/1gluon_jetfunc_low} \otimes \fd{3cm}{figures_b/eikonal_factor_gluon_low.pdf}  }^{\text{Collinear Gluon Correction}}   \nn \\
&= \int\!\! d\omega_2\: \Re\Bigl[C^{(2)}_{\cP \cB} C^{(0)}\Bigr](\omega_2)\, J^{(2)}_{\cB \cP}(\omega_2) \otimes J_{g,\bn}^{(0)} \otimes S_g^{(0)} + n \leftrightarrow \bn\,,
\end{align}
where $S_{\bn B_{us}}^{(2)}$ is the subleading soft function involving an explicit gauge invariant gluon field, while $J^{(2)}_{\cB \cP}$ is a subleading power jet function which arises from hard scattering operators involving an additional $B_\perp$ field, and $\cP_\perp$ operator. Those are naturally derived from the hard scattering operators in \Tab{tab:higgsops}.
More details about those functions can be found in \cite{Moult:2018jjd}. No radiative functions contribute at leading log for this category due to helicity selection rules \cite{Moult:2018jjd,Moult:2019mog}.

In category 2, i.e.~for the quark corrections, we have a soft quark radiative function contribution
\begin{align}
& \overbrace{  \left| \fd{1.5cm}{figures_new/2gluon_hard_low.pdf}  ~\right|^2 \cdot   \int dr_2^+ dr_3^+ \fd{3cm}{figures_b/soft_quark_collinear_piece_gluoncase_purjet_low.pdf}\otimes \fd{3cm}{figures_b/1gluon_jetfunc_low} \otimes  \fd{3cm}{figures_b/soft_quark_diagram_wilsonframe_gluoncase_low.pdf} }^{\text{Soft Quark Correction}}\nn \\
&= \big|C^{(0)}\big|^2 \int dr_2^+ dr_3^+ \: J_{n \psi}^{(2)}(r_2^+,r_3^+) \otimes J_{g,\bn}^{(0)} \otimes S_{\psi}^{(2)}(r_2^+,r_3^+) + n \leftrightarrow \bn\,,
\end{align}
and a subleading hard scattering operator contribution
\begin{align}
& \overbrace{ \int d \omega_1 d \omega_2  \left| \fd{1.5cm}{figures_new/collinear_category2_hard_low}~ \right|^2  \otimes\fd{3cm}{figures_b/1quark1gluon_jetfunc_low}\otimes \fd{3cm}{figures_b/1quark_jetfunction_low}\otimes \fd{3cm}{figures_b/eikonal_factor_gluon_low.pdf} }^{\text{Collinear Quark Correction}}\nn\\
&= \int\!\! d\omega_a d \omega_b\: \Re\Bigl[C^{(1)}_{\cB n} C^{(1)}_{\cB \bn}\Bigr](\omega_a,\omega_b)\, J^{(2)}_{\cB \cB}(\omega_a,\omega_b) \otimes J_{q,\bn}^{(0)} \otimes S_g^{(0)} + n \leftrightarrow \bn
\end{align}
where $J^{(2)}_{\cB\cB}$ is a four gluon subleading power jet function arising from the hard scattering operators $\cO^{(1)}_{\cB n}$, and $\cC^{(1)}_{\cB n}$ is the Wilson coefficient of $\cO^{(1)}_{\cB n}$. Note that this operator and Wilson coefficient are the same ones that appear in \eq{Higgs_match} and that, in full QCD, their corresponding contributions give rise to the endpoint behavior discussed in \sec{factorization}. The definitions of the radiative functions $J_{n \psi}$, $S_{\psi}$ are given in \Tab{tab:radHiggs}.

\subsection{$e^+e^-$ annihilation}\label{app:app_factorization_ee}
We now repeat the same exercise for the case of the ingredients entering the leading log cross section for thrust in $e^+e^-$ annihilation at next-to-leading power.
In \Tab{tab:eeops} we list the hard scattering operators contributing at leading logarithm for this process.

{
\renewcommand{\arraystretch}{1.4}
\begin{table}[h]
\begin{center}
\scalebox{0.842}{
\begin{tabular}{| l | c | c |c |c|c| r| }
  \hline                       
   Operator & Tree Level Matching Coefficient \\
  \hline
    $\cO^{(0)}=C^{(0)} \bar\chi_n \Gamma \chi_\bn$& $C^{(0)}=1$ \\
  \hline
  $ \cO^{(1)\mu}_{\chi\chi\bar n1}= C^{(1)}_{\chi\chi\bar n1}\bar \chi_{\bar n, \omega_1} \Sl{\cB}_{\perp  n,\omega_3} \frac{\Sl{n}}{2}  \Gamma  \chi_{\bar n,-\omega_2}$ & $C^{(1)}_{\chi\chi\bar n1}=-\frac{g}{\omega_1}  $ \\
  \hline
  $ \cO^{(1)\mu}_{\chi\chi\bar n2}= C^{(1)}_{\chi\chi\bar n2} \bar \chi_{\bar n, \omega_1}   \Gamma \frac{\Sl{n}}{2} \Sl{\cB}_{\perp n,\omega_3}   \chi_{\bar n,-\omega_2}$ & $C^{(1)}_{\chi\chi\bar n2}=\frac{g}{\omega_2}  $ \\
  \hline
   $\mathcal{O}^{(2)\mu}_{\cP \bar n1} = C^{(2)}_{\cP \bar n1} \bar \chi_{ n,\omega_1} \big[\Sl\cB_{\perp \bar n, \omega_3} \Sl{\cP}^\dagger_{\perp}\big]  
  \Gamma \chi_{\bar n,-\omega_2}$ &  $C^{(2)}_{\cP \bar n1}=- \frac{g}{\omega_1 \omega_3} $ \\
   \hline
	$\mathcal{O}^{(2)\mu}_{\cP \bar n2} = C^{(2)}_{\cP \bar n2} \bar \chi_{n,\omega_1}  \Big[\Sl\cB_{\perp \bar n, \omega_3} \frac{\Sl{\bar n}}{2} \Gamma \frac{\Sl{n}}{2} \Sl{\cP}^\dagger_{\perp}\Big]  \chi_{\bar n,-\omega_2}$  & $C^{(2)}_{\cP \bar n2}=- \frac{g}{\omega_1 \omega_2} $\\
  \hline  
\end{tabular}}
\end{center}
\caption{
Hard scattering operators that contribute to the thrust cross section to $\cO(\lambda^2)$ at LL, along with their tree level matching coefficients. These operators and matching coefficients were derived in \cite{Feige:2017zci}. Note that, for the case of $e^+e^-$ annihilation the Wilson coefficients of the hard scattering operators involving ultrasoft emissions, vanish at tree level, and therefore do not contribute at LL~\cite{Feige:2017zci}. For a comparison between the subleading hard scattering operators for the case of $e^+e^-$ annihilation and gluon induced Higgs decay, see Sec. 3.4.3 of \cite{Moult:2017rpl}.
}
\label{tab:eeops}
\end{table}
}

{
\renewcommand{\arraystretch}{1.4}
\begin{table}[h]
\scalebox{0.842}{
\hspace{0.2cm}\begin{tabular}{| l | c | c |c |c|c| r| }
  \hline                       
  $T$-Product & Example Diagram & Soft Function& Jet Function \\
  \hline
  $\cL^{(2)}$ & $\fd{3cm}{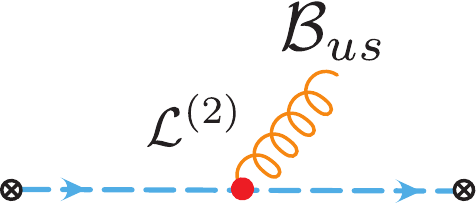}$ &  $\langle0| Y_n  Y_{\bar n}(0) \partial \cB_{us(n)}(x)Y_{\bar n} Y_n(0)   |0\rangle$& $ \langle0| \bar \chi_n(y) \bar \chi_n \chi_n(r_2^+)  \chi_n (0)|0\rangle$ \\
  \hline
   $\cL^{(1)}\cdot \cL^{(1)}$ & $\fd{3cm}{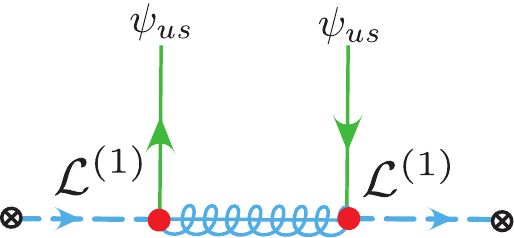}$ & {\begin{small}$ \langle0|Y_n  Y_{\bar n}(0) \bar \psi_{us}(x) \psi_{us}(z)Y_{\bar n} Y_n(0)  |0\rangle$\end{small}} &  {\begin{small}$ \langle0| \bar \chi_n(y) \bar \chi_n \cB_{n\perp} (z)\cdot  \bar \chi_n \cB_{n\perp} (x) \chi_n (0)|0\rangle$  \end{small}} \\
  \hline  
\end{tabular}}
\caption{
Leading order contributions to the $\cO(\lambda^2)$ thrust cross section from radiative functions. The soft gluon emission arises from a single $\cL^{(2)}$ insertions, in accord with the LBK theorem, while the soft quark contribution arises from two $\cL^{(1)}$ insertions. We do not explicitly write the corresponding functions when the soft gluon or quark is emitted from the $\bar n$ collinear sector. The results in this table have been obtained in \cite{Moult:2019mog}
}
\label{tab:radee}
\end{table}
}

For category 1, we have a contribution from a radiative function
\begin{align}\label{eq:softgluon_ee}
&\overbrace{ \left| \fd{1.5cm}{figures_b/2quark_hardfunc_low.pdf} ~ \right|^2\cdot\int d r_2^+ \fd{3cm}{figures_b/radiative_quarkcase_low.pdf} \otimes \fd{3cm}{figures_b/1quark_jetfunction_low}\otimes  \fd{3cm}{figures_b/wilson_framing_radiative_low.pdf} }^{\text{Soft Gluon Correction}}\nn\\
&= \big|C^{(0)}\big|^2 \int dr_2^+ \: J_{\cB_{us}}^{(2)}(r_2^+) \otimes J_{q,\bn}^{(0)} \otimes S_{\cB_{us}}^{(2)}(r_2^+) + n \leftrightarrow \bn\,,
\end{align}
and one from a subleading hard scattering operator
\begin{align}
&\overbrace{ \int d\omega_1 \Re\left ( \fd{1.5cm}{figures_b/collinear_category1_hard_low}\cdot \fd{1.5cm}{figures_b/2quark_hardfunc_low.pdf}^\dagger \right )  \otimes\fd{3cm}{figures_b/gq_jetfunction_low}\otimes \fd{3cm}{figures_b/1quark_jetfunction_low} \otimes \fd{3cm}{figures_b/eikonal_factor_gluon_low.pdf}}^{\text{Collinear Gluon Correction}}\nn \\
&= \int\!\! d\omega_1\: \Re[C^{(2)}_{\cP \bar n2} C^{(0)} ](\omega_1)\, J^{(2)}_{\cB \cP}(\omega_1) \otimes J_{q,\bn}^{(0)} \otimes S_q^{(0)} + n \leftrightarrow \bn\,,
\end{align}
where $J^{(2)}_{\cB \cP}$ is a subleading power jet function which arises from hard scattering operators involving an additional $B_\perp$ field, and $\cP_\perp$ operator and $S_{\cB_{us}}^{(2)}$ is a subleading soft function involving an insertion of gauge invariant gluon field along the lightcone. More details on this can be found in Section 6.2.2 of \cite{Moult:2019mog}.

For category 2, i.e.~the quark corrections, we have one contribution from a radiative soft function
\begin{align}
&\overbrace{   \left| \fd{1.5cm}{figures_b/2quark_hardfunc_low.pdf}~ \right|^2  \cdot   \int dr_2^+ dr_3^+ \fd{3cm}{figures_b/soft_quark_collinear_piece_purjet_low.pdf}\otimes \fd{3cm}{figures_b/1quark_jetfunction_low} \otimes  \fd{3cm}{figures_b/soft_quark_diagram_wilsonframe_low.pdf}   }^{\text{Soft Quark Correction}}\nn \\
&= \big|C^{(0)}\big|^2 \int dr_2^+ dr_3^+ J_{n \psi}^{(2)}(r_2^+,r_3^+)\otimes J_{q,\bn}^{(0)} \otimes S^{(2)}_{\psi}(r_2^+,r_3^+) + n \leftrightarrow \bn
\,,
\end{align}
and one from the product of two $\cO(\lambda)$ hard scattering operators
\begin{align}
&\overbrace{  \int d \omega_1 d \omega_2 \left| \fd{1.5cm}{figures_b/collinear_category2_hard_low}~ \right|^2\otimes\fd{3cm}{figures_b/2quark_jetfunc_low}\otimes \fd{3cm}{figures_b/1gluon_jetfunc_low}\otimes \fd{3cm}{figures_b/eikonal_factor_gluon_low.pdf} }^{\text{Collinear Quark Correction}} \nn \\
&=\int d\omega_1 \omega_2 \Big|C^{(1)}_{\chi \chi n}(\omega_1) C^{(1)}_{\chi \chi n}(\omega_2) \Big| J^{(2)}_{\chi \chi}(\omega_1,\omega_2) \otimes J_{g,\bn}^{(0)} \otimes S_q^{(0)} + n \leftrightarrow \bn\,.
\end{align}
Here $J^{(2)}_{\chi \chi}$ is a four quark subleading power jet function arising from the hard scattering operators $\cO^{(1)}_{\chi \chi n}$, whose Wilson coefficient $\cC^{(1)}_{\chi \chi n} (\omega)$ is quoted in \Tab{tab:eeops}, and shows a singular end point behavior as $\omega\to 0$. Note that this operator and Wilson coefficient are the same that appear in \eq{matching_qqsame1} and that, in full QCD, the corresponding contributions give rise to the endpoint behavior discussed in \sec{factorization}. The definitions of the radiative functions $J_{n \psi}$, $S_{\psi}$ are given in \Tab{tab:radee} and the interested reader can find a detailed derivation of these contributions in Section 6.2.1 of \cite{Moult:2019mog}.

\bibliography{subRGE}{}
\bibliographystyle{jhep}

\end{document}